\documentclass[11pt,fleqn]{article}
\usepackage{graphicx,enumerate}
\usepackage{amsmath}
\usepackage{amssymb}
\usepackage{color}
\usepackage{hyperref}
\hypersetup{colorlinks = true, linktocpage = true, bookmarksopen = true, linkcolor = blue, urlcolor=blue, citecolor = blue, allcolors=blue}
\usepackage[caption=false,position=top]{subfig}
\usepackage{authblk}
\usepackage[font=footnotesize]{caption}
\usepackage[numbers,sort&compress]{natbib}
\numberwithin{equation}{section}
\usepackage[margin=2.5cm]{geometry}
\newcommand{\ba}{\begin{array}}
\newcommand{\ea}{\end{array}}
\newcommand{\be}{\begin{equation}}
\newcommand{\ee}{\end{equation}}
\newcommand{\bc}{\begin{center}}
\newcommand{\ec}{\end{center}}
\newcommand{\bdm}{\begin{displaymath}}
\newcommand{\edm}{\end{displaymath}}

\newcommand{\p}{\partial}

\newcommand{\bfm}[1]{\mbox{\boldmath $ #1 $}}

\usepackage{array}
\newcolumntype{P}[1]{>{\raggedleft\arraybackslash}p{#1}}

\begin{document}
\title{\bf Modelling drug delivery from multiple emulsions} 

\author[1]{G.~Pontrelli\footnote{Corresponding author: \href{mailto:giuseppe.pontrelli@gmail.com}{giuseppe.pontrelli@gmail.com}.}}
\author[2]{E. J.~Carr}
\author[1,3]{A. Tiribocchi}
\author[1,3]{S. Succi}

\affil[1]{{\footnotesize Istituto per le Applicazioni del Calcolo -- CNR
Via dei Taurini 19 -- 00185 Rome, Italy}}
\affil[2]{{\footnotesize School of Mathematical Sciences, Queensland University of Technology (QUT), Brisbane, Australia}}
\affil[3]{{\footnotesize Italian Institute of Technology, CNLS@Sapienza, Rome, Italy}}

\maketitle
\begin{abstract}
We present a mechanistic model of drug release from a multiple emulsion into an external surrounding fluid. We consider a single multi-layer droplet where the drug kinetics are described by a pure diffusive process through different liquid shells. The multi-layer problem is described by a system of diffusion equations coupled via interlayer conditions imposing continuity of drug concentration and flux. Mass resistance is imposed at the outer boundary through the application of a surfactant at the external surface of the droplet. The two-dimensional problem is solved numerically by finite volume discretization. Concentration profiles and drug release curves are presented for three typical round-shaped (circle, ellipse and bullet) droplets and the dependency of the solution on the mass transfer coefficient at the surface analyzed. The main result shows a reduced
release time for an increased elongation of the droplets. 
\end{abstract}
\vspace*{2ex}\noindent\textit{\bf Keywords}: droplets, nanoemulsions, drug release, multi-layer diffusion, numerical solutions.


\section{Introduction}

Multiple emulsions consist of a dispersion of immiscible spherical fluid droplets, of diameter ranging from $1$ to $50$ $\mu$m, in a larger fluid drop, of size up to $100$$\mu$m \cite{garti1998,utada_2015,weitz,vladi_2017}. The simplest low-ordered realization is the double emulsion, where, for instance, a water core is surrounded by a thin concentric oil layer. If the double emulsion is immersed in water, it is often termed as a water/oil/water (W/O/W) emulsion. More complex examples include collections of polydisperse droplets placed in a larger drop or multi-layer distinct cores of fluid \cite{ding2019,vladi_2017}. Their stability is generally guaranteed by a surfactant (adsorbed onto the external interface) which prevents coalescence of the droplets \cite{utada_2015,weitz,weitz2,weitz3}.
Such emulsions are conventionally manufactured by means of microfluidic devices, which, alongside a large production rate, ensure a high degree of reproducibility \cite{vladi_2017}. Due to their compartmental structure, these systems are extensively used to encapsulate and transport active components in a number of technological applications, including food processing \cite{weiss,calleros}, cosmetics \cite{yoshida,lee}, syntheses of microspheres and microcapsules \cite{utada_2015,lorenceau,bocanegra,rizk}, to name a few. 

Multi-layered emulsions are particularly suited as drug carriers of pharmaceutical and biological compounds, due to their capability to combine an efficient mechanical stability to a controlled release of the cargo within the range of the therapeutic window \cite{pays_2002,qi_2011}. Indeed, unlike a layer-free emulsion, the multi-layer assembly ensures protection of the active agent against external chemical aggression as well as an enhanced control of the transfer rate by the thin oil barrier \cite{koker,timin}. These features drastically diminish the premature degradation of the compound and broaden the sustainability of the emulsion. A further benefit stems from its inherent soft structure, which can be selectively hardened or gelled by tuning the viscosity of the middle fluid layer \cite{omi_2003,chu_2003}. In contrast to rigid capsules, this allows for, for instance, migration through narrow interstices where large shape deformations are expected to occur.


Although many efforts have been dedicated to the experimental realization of optimized drug-delivery via multiple emulsions, much less is known about the underlying mechanism governing the drug release in these systems. In pharmaceutical applications, the drug is usually stored within the internal water droplet and then, after diffusing through the surrounding oil shell, is subsequently released in the external medium. Amongst several physico-chemical processes, such as osmosis and drug dissolution, diffusion is by far the dominant mechanism controlling drug kinetics and  release \cite{grassi}. This process is crucially influenced by the medium properties as well as by the ultra-thin surfactant layer confined at the droplet interface. Indeed, the latter may partially hinder the mass flux of the drug towards the external medium and, hence, potentially compromise its efficacy \cite{pays_2002}.

\begin{figure}[t]
\centering
\subfloat[ \hspace{0.23\textwidth} (b) \hspace{0.25\textwidth} (c)]
{\includegraphics[scale=0.8]{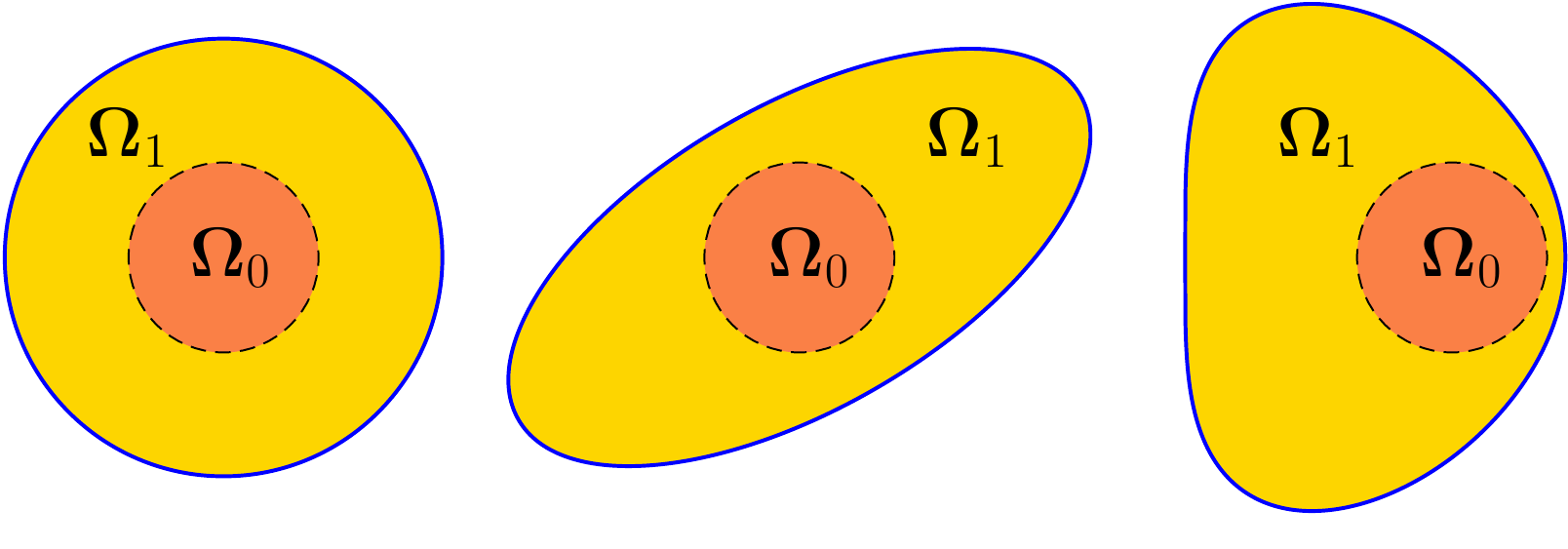}}
\caption{Schematic representation of the cross-section of the droplet comprised of an internal circular core $\Omega_0$, and an enveloping denser fluid shell $\Omega_1$. A thin membrane (shown in blue) is present at the surface modelling the surfactant finite resistance.}  
\label{domain1}
\end{figure}

In many practical situations, such as a capsule migrating in a blood vessel, emulsions are dragged by the surrounding fluid. Even under weak shears (those typical of a laminar regime in a microfluidic channel), the flow is known to produce relevant shape deformations that may potentially alter the functioning of the multi-core emulsion as a drug carrier \cite{chen,chen2,prl_cruz,wang_2013prl,tiribocchi}. Under a mild steady extensional flow, for instance, a spherical-shaped double emulsion may turn into an ellipsoid \cite{chen,chen2,prl_cruz}, whereas bean or bullet-like shapes emerge when the emulsion is subject to a Poiseuille flow \cite{wang2,nabavi,guido}, leaving the shape of the inner core essentially unaffected. More complex effects are observed in the presence of more intense flows, such as an iper-stretching of the core in tandem with the outer drop leading to their breakup and formation of two daughter cores \cite{prl_cruz,basa}. Hence, it is of particular relevance to understand how the geometry of the emulsion can influence the drug transport. In this respect, the development of semi-empirical and mechanistic models is crucial for predicting the release performances in multiple emulsions and for improving their design. Besides providing a systematic approach to solve these tasks, mathematical modeling can also serve as a tool to answer practical issues, such as the identification of the parameters to tune in order to achieve a predetermined delivery rate or the development of physico-chemical markers capturing main transport processes \cite{siep, pep}. While previous studies have been focused on modeling drug release in multi-layered rigid spherical microcapsules \cite{timin,kaoui,carr}, in this work we extend the mechanistic approach to soft multiple emulsions focussing on the shapes shown in Fig.~\ref{domain1}, which are geometries  observed at equilibrium or with a very weak flow (a), and under extensional/shear (b), and Poiseuille flow (c), in the laminar regime (i.e. when the Reynolds number remains below $1$).

We describe the drug kinetics from a double emulsion by means of a system of diffusion equations coupled via suitable boundary and interlayer conditions. We simulate the transport of the drug initially confined within the inner spherical core and compute its release time, by varying the diffusivity of the shell and the permeability of the external membrane due to a surfactant.
Our results show how the geometry of the double emulsion does have an influence on the drug delivery and, in particular, how elongated droplets exhibit a faster release.
Once the parameters are identified, the proposed methodology
provides a simple tool that can be used to quantitatively characterize
the drug transport, improve the technological performance and
optimize the release rate for therapeutic purposes.  
The remaining sections of this paper are structured as follows. In Sect.~\ref{sec:model} we describe the equations governing the kinetics of the drug in a core-shell emulsion geometry and in Sect.~\ref{sec:numerical} we illustrate the details of the numerical model involving a finite volume discretisation over an unstructured mesh. Sect.~\ref{sec:results} is devoted to the presentation and discussion of numerical results of a drug releasing round-shaped droplet, under
different flow conditions.



\section{Drug diffusion from a multi-layer droplet}
\label{sec:model}

\begin{figure}[t!]
\centering
\includegraphics[scale=0.7]{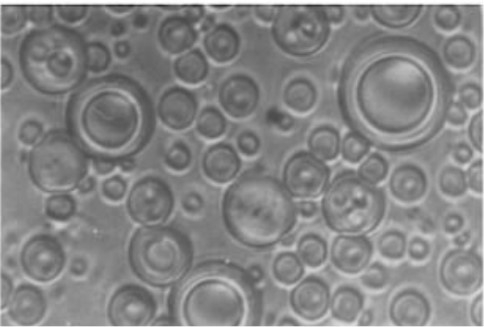}
\caption{A snapshot of a typical double emulsion (from \cite{ding2019}).}  
\label{fig2}
\end{figure}

In the most general case, a composite emulsion-based droplet is comprised of $n$ enveloping concentric liquid shells. This multi-layer droplet is immersed in an external release medium (or bulk fluid) that, for the purposes of this work, is assumed  stable and steady. Without loss of generality and with reference to Fig.~\ref{fig2}, we restrict our analysis to a single vesicle, utilizing the superposition principle for the release from a number of identical droplets. Among several possible double emulsion configurations \cite{ding2019}, we consider here a droplet $\Omega$ constituted of two concentric layers of fluid, a circular core $\Omega_0$ (layer 0) and an enveloping shell $\Omega_1$ (layer 1). As the emulsion is fabricated through a confined fluid flow, the external layer is typically deformed in the flow direction. Neglecting droplet microfluidics and deformation, we restrict our attention to the drug delivery and the characteristics of the release for a fixed shape. In actual fact, with the superimposed flow at steady state, droplets assume and maintain a variety of typical shapes, from the round and oblate spheroid, to an ellipsoid or bullet-like geometry, with the spherical shape of the internal core preserved (see Fig.~\ref{domain1}). 
In this study we consider a 2D cross section of the  droplet aligned with the superimposed two-dimensional fluid flow.
Generally a surfactant is added to the surface of  the vesicle to prevent coalescence \cite{pays_2002}, and this results in a additional resistance to the drug release. To include this effect, a thin membrane is assumed at the surface of the droplet with a surface mass transfer coefficient $P$ (m/s) expressing the surfactant finite resistance (Fig.~\ref{domain1}) \cite{kaoui, carr}.

In a steady and stable double emulsion, we assume the drug kinetics are governed by a purely diffusive two-layer model, where the evolution of the concentrations, $c_0({\bfm x},t)$ and $c_1({\bfm x},t)$, in the core and shell respectively, are governed by a set of 2D linear diffusion equations \cite{carr,kaoui}:
\begin{align}
&{\p c_0 \over \p t} =\nabla\cdot({D_0 \nabla c_0}),\qquad {\bfm x}\in\Omega_{0},\label{eq:model_pde1}  \\
&{\p c_1 \over \p t} =\nabla\cdot({D_1 \nabla c_1}),\qquad {\bfm x}\in\Omega_{1}, \label{eq:model_pde2} 
\end{align}
paired with the following interlayer, boundary and initial conditions
\begin{alignat}{2}
&c_0 = c_1, \quad D_0 { \nabla c_0}\cdot\boldsymbol{n}_{\Gamma} = D_1 { \nabla c_1 }\cdot\boldsymbol{n}_{\Gamma}, &\qquad& {\bfm x}\in\Gamma, \\
& D_1{\nabla c_1}\cdot\boldsymbol{n}_{\Omega} =  -P c_1, &\qquad& {\bfm x}\in\p\Omega, \label{eq:model_int2}\\
&c_0({\bfm x},0)=C_0, &\qquad& {\bfm x}\in \Omega_0, \label{eq:model_ic1}\\
&c_1({\bfm x},0)=C_1, &\qquad& {\bfm x}\in \Omega_1,  \label{eq:model_ic2}
\end{alignat}
where $C_0, C_1>0$ are constants, $\Gamma$ is the interface between $\Omega_{0}$ and $\Omega_{1}$, $\boldsymbol{n}_{\Gamma}$ is a unit normal to $\Gamma$ and $\boldsymbol{n}_{\Omega}$ is the unit normal to $\partial\Omega$ directed outwards from $\Omega$ . In the above equations, the parameters $D_0, D_1$ are the drug diffusion coefficients of the two layers and $P$ is the specified mass transfer coefficient at the surface \cite{kaoui,carr}. In the limit $P \rightarrow 0$, we have an impermeable membrane, when  $P \rightarrow \infty$ we recover a perfect sink condition (no resistance).

All the variables, the parameters and the equations are scaled by means of the change of variables:\be
\label{eq:nondim_constants}
{\bfm x} \rightarrow {{\bfm x} \over \chi}, \quad  t \rightarrow { D_{\max} \over \chi^2} \, t , \quad  c_i  \rightarrow {c_i \over C_{\mathrm{max}}},   \quad  \text{$i=0,1$},
\ee
and by redefining the non-dimensional constants:
\be
  D_i \rightarrow {D_i \over  D_{\max}},   \quad  C_i  \rightarrow {C_i \over C_{\mathrm{max}}},  \quad  P \rightarrow {P \, \chi  \over D_{\mathrm{max}}}, \quad \text{$i=0,1$}, \label{gh6}
\ee
where $\chi$ is a characteristic length scale of $\Omega_{1}$, $C_{\max} = \max ({C_0, C_1})$ and $D_{\max} = \max({D_0, D_1})$. 
\section{Numerical method}
\label{sec:numerical}

\subsection{Solving for the drug concentration}
\label{sec:numerical_concentration}
The non-dimensionalized analogue of the diffusion model (\ref{eq:model_pde1})--(\ref{eq:model_ic2}) is solved numerically by discretizing in space using a finite volume method on an unstructured mesh (see, e.g., \cite{fvm}). To perform the meshing, we use the mesh generator GMSH \cite{gmsh} to construct meshes consisting of a set triangular elements ($T_{\Omega}$). Each element is located entirely within either $\Omega_{0}$ or $\Omega_{1}$ (i.e., elements adjacent to the interface, $\Gamma$, have an edge that aligns with the interface) with $T_{\Omega_{0}}$ and $T_{\Omega_{1}}$ used to denote the set of elements located in $\Omega_{0}$ and $\Omega_{1}$, respectively. 

\begin{figure}[t]
\centering
\includegraphics[width=0.5\textwidth]{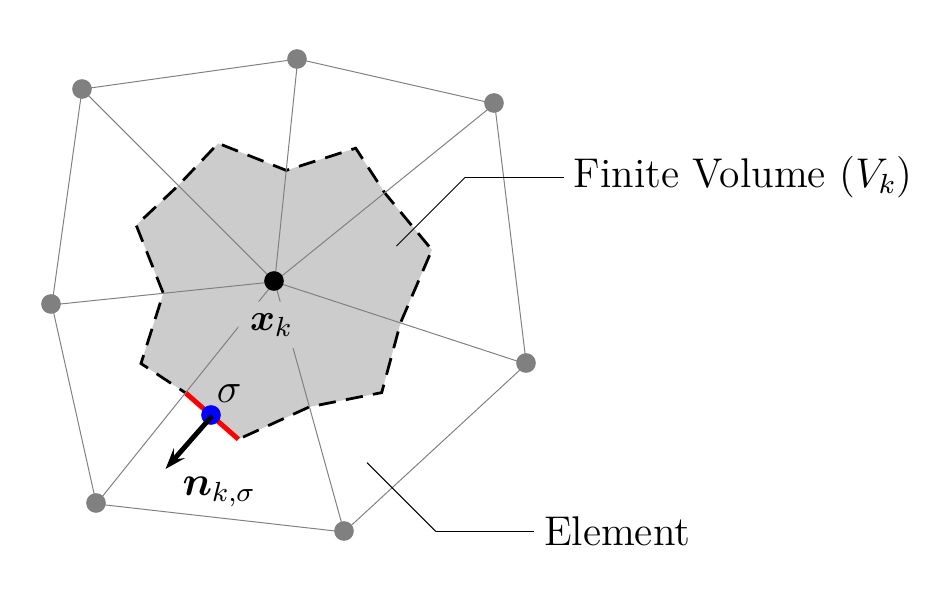}
\caption{Notation used in the finite volume discretization. Depicted is the finite volume ($V_{k}$) corresponding to an arbitrary internal node $k$ with the boundary of the finite volume shown using a dashed line. The blue dot locates the midpoint of edge $\sigma$ ($\boldsymbol{x}_{\sigma}$) and the length of the red edge, labelled $\sigma$, is $L_{\sigma}$.}
\label{fig:fvm}
\end{figure}

We employ a vertex-centered strategy, where finite volumes are constructed around each node by connecting the centroid of each triangular element to the midpoint of its edges (Fig.~\ref{fig:fvm}). Spatial discretization is applied to the following equivalent form of (\ref{eq:model_pde1})--(\ref{eq:model_pde2}):
\begin{align*}
\frac{\partial c}{\partial t} = \nabla\cdot\left(D(\boldsymbol{x})\nabla c\right),\qquad\text{$\boldsymbol{x}\in\Omega$},
\end{align*}
where 
\begin{align*}
c(\boldsymbol{x},t) = \begin{cases} c_{0}(\boldsymbol{x},t), & \text{if $\boldsymbol{x}\in\Omega_{0}$},\\
c_{1}(\boldsymbol{x},t), & \text{if $\boldsymbol{x}\in\Omega_{1}$},\end{cases}\qquad
D(\boldsymbol{x}) = \begin{cases} D_{0}, & \text{if $\boldsymbol{x}\in\Omega_{0}$},\\
D_{1}, & \text{if $\boldsymbol{x}\in\Omega_{1}$}.\end{cases}
\end{align*}
Let $N$ be the number of nodes in the mesh, $\widetilde{c}_{k}:=\widetilde{c}_{k}(t)$  be the numerical approximation to $c(\boldsymbol{x}_{k},t)$ and $V_{k}$ be the finite volume surrounding node $k$ for all $k = 1,\hdots,N$ (Fig.~\ref{fig:fvm}). The finite volume discretization yields the follow system of spatially-discrete equations: 
\begin{align}
\label{eq:fvm}
\frac{\text{d}\widetilde{c}_{k}}{\text{d}t} = \frac{1}{|V_{k}|}\sum_{\sigma\in \mathcal{E}_{k}} F_{k,\sigma},\qquad k = 1,\hdots,N,
\end{align}
where $\mathcal{E}_{k}$ is the set of edges comprising the boundary of $V_{k}$, $|V_{k}|$ is the area of $V_{k}$ and $F_{k,\sigma}$ is a numerical approximation to the (negative) flux $\int_{\sigma}D(\boldsymbol{x})\nabla c\cdot\boldsymbol{n}_{k,\sigma}\,\text{d}\boldsymbol{x}$ with $\boldsymbol{n}_{k,\sigma}$ denoting the unit vector normal to edge $\sigma$ directed outward from $V_{k}$ (Fig.~\ref{fig:fvm}). The value of $F_{k,\sigma}$ depends on whether the edge $\sigma$ is located in the interior of the droplet ($\Omega$) or along the boundary ($\partial\Omega$):
\begin{align}
\label{eq:flux}
F_{k,\sigma} = \begin{cases} [D(\boldsymbol{x}_{\sigma})(\widetilde{\nabla c})_{\sigma}\cdot\boldsymbol{n}_{k,\sigma}]L_{\sigma}, & \text{if $\boldsymbol{x}_{\sigma}\in\Omega$},\\[0.2cm]
-P \widetilde{c}_{\sigma} L_{\sigma}, & \text{if $\boldsymbol{x}_{\sigma}\in\partial\Omega$}, \end{cases}
\end{align}
with $\boldsymbol{x}_{\sigma}$ and $L_{\sigma}$ denoting the midpoint and length of edge $\sigma$, respectively. The quantities $\widetilde{c}_{\sigma}$ and $(\widetilde{\nabla c})_{\sigma}$ are numerical approximations to $c(\boldsymbol{x}_{\sigma},t)$ and $\nabla c(\boldsymbol{x}_{\sigma},t)$ computed/discretised by assuming the concentration varies linearly within each triangular element. The discretised forms for $\widetilde{c}_{\sigma}$ and $(\widetilde{\nabla c})_{\sigma}$ are expressed in terms of $\widetilde{c}_{k}$ for $k\in\mathcal{N}_{\sigma}$, where $\mathcal{N}_{\sigma}$ is the set of three nodes corresponding to the three vertices of the triangular element in which $\sigma$ is located. In summary, the finite volume equations (\ref{eq:fvm})--(\ref{eq:flux}) define a system of linear ordinary differential equations, expressible in matrix form as:
\begin{align}
\label{eq:ode_system}
\frac{\text{d}\mathbf{c}}{\text{d}t} = \mathbf{A}\mathbf{c},\qquad \mathbf{c}(0) = \mathbf{c}_{0},
\end{align}
where $\mathbf{c} = (\widetilde{c}_{1},\hdots,\widetilde{c}_{N})^{T}$, $\mathbf{A}$ is an $N\times N$ matrix and $\mathbf{c}_{0}$ is the discretised form of the initial conditions (\ref{eq:model_ic1})--(\ref{eq:model_ic2}) with the $k$th entry of $\mathbf{c}_{0}$ equal to $C_{0}$ if $\boldsymbol{x}_{k}\in\Omega_{0}$, $C_{1}$ if $\boldsymbol{x}_{k}\in\Omega_{1}$ and the weighted average $(|V_{k}\cap\Omega_{0}|C_{0}+|V_{k}\cap\Omega_{1}|C_{1})/|V_{k}|$ if $\boldsymbol{x}_{k}\in\Gamma$. The system (\ref{eq:ode_system}) is solved using MATLAB's built-in \texttt{ode15s} solver with the default options and tolerances \cite{ode15s}.

\subsection{Computing the drug mass}
The drug mass in the droplet layers (core and shell) and the total drug mass are defined as follows:
\be
\label{eq:M0_M1}
M_0(t)= \int_{\Omega_0}  c_0({\bfm x},t)\,\mathrm{d}{\bfm x}, \qquad M_1(t)=  \int_{\Omega_1}  c_1({\bfm x},t)\,\mathrm{d}{\bfm x},\qquad M_{T}(t) = M_{0}(t) + M_{1}(t).   
\ee
These quantities are calculated from the numerical solution described in Sect.~\ref{sec:numerical_concentration} by integrating the piecewise linear concentration across each element yielding the approximations: 
\begin{gather}
M_{0}(t) \approx \sum_{E\in T_{\Omega_{0}}} \text{mean}\{\widetilde{c}_{k}(t)\,|\,k\in \mathcal{N}_{E}\}\cdot|E|,\qquad M_{1}(t) \approx \sum_{E\in T_{\Omega_{1}}} \text{mean}\{\widetilde{c}_{k}(t)\,|\,k\in\mathcal{N}_{E}\}\cdot|E|,\\
M_{T}(t) \approx \sum_{E\in T_{\Omega}} \text{mean}\{\widetilde{c}_{k}(t)\,|\,k\in\mathcal{N}_{E}\}\cdot|E|,
\end{gather}
where $\mathcal{N}_{E}$ is the set of three nodes corresponding to the three vertices of triangular element $E$ and $|E|$ is the area of element $E$. With the masses in both layers calculated, the \textit{fractional released mass}, i.e. the fraction of the initial mass that has been released at time $t$, is computed as
\be
\label{eq:fractional_released_mass}
M_r(t)= 1-{M_{T}(t) \over M_{T}(0)}.
\ee
Note that $M_r(0) = 0$ and $\lim\limits_{t \rightarrow \infty} M_r(t) = 1$ since $\lim\limits_{t \rightarrow \infty} M_0(t) = M_1(t)=0$. The release time, defined as the time $t^*$ at which $ M_r(t^*) \approx 1$, is specified in the next section. 

\section{Results and discussion}
\label{sec:results}
Among a variety of configurations, we want to analyze the sensitivity of the release with respect to the properties of the shell $\Omega_1$ for varying mass transfer coefficient $P$ and diffusion coefficient $D_1$, when the core $\Omega_0$ and the initial drug mass are kept unchanged. The parameters used are consistent with typical values in the literature and listed in Table \ref{tab:parameters}.

\begin{table}[h!]
\renewcommand*{\arraystretch}{1.1}
  \begin{center}
\caption{Nondimensional range and value of the parameters}
 \label{tab:parameters}
  \begin{tabular}{|c|c|c|c|}
	               \hline
                 {Model parameters} & {Physical range} &  {Simulated values}  & {References} \\
                \hline
                $R_0$ ($\mu$m) &  $0.5-50$ & 40  &\cite{utada_2015,dluska_2016,pays_2002,vladi_2017,schwarz_2012}\\
                \hline
                $R_{1}$ ($\mu$m) &  $50-100$ &  50, 80  &\cite{zhang_2013_bio,utada_2015,pays_2002,vladi_2017,schwarz_2012}\\
                \hline
                 $D_0$ (m$^2$/s)  &  $10^{-9}$-$10^{-10}$ &  $10^{-10}$ & \cite{mezzenga_2004}\\
                \hline
                 $D_1$ (m$^2$/s) &   $10^{-11}$-$10^{-13}$ &  $10^{-12},10^{-13}$  &\cite{mezzenga_2004}\\
                \hline
                $P$ (m/s)  &   0-1 &  $10^{-7}, 2 \cdot 10^{-4}$  &  \cite{zhang_2013_bio,tewes_2007,chan_2013}  \\
								\hline
\end{tabular}
 \end{center}
\end{table} 

\begin{figure}[t]
\centering
\includegraphics[width=0.98\textwidth]{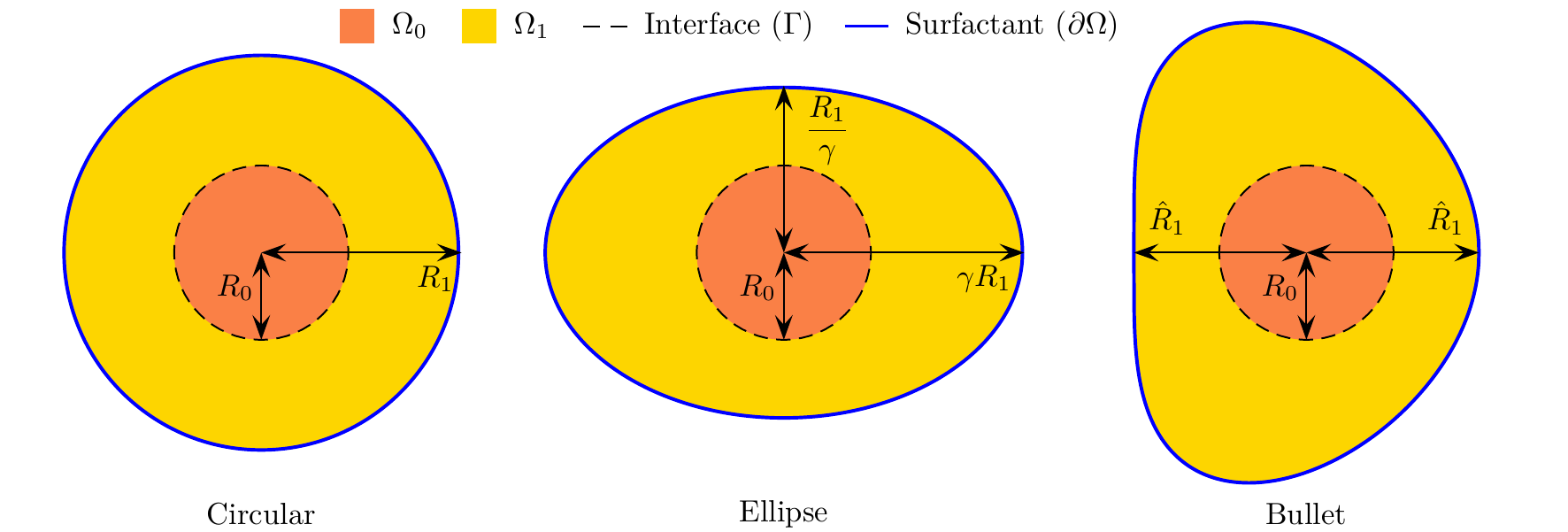}
\caption{Schematic representation of the cross-section of the droplet comprised of an internal circular core $\Omega_0$, and an enveloping shell $\Omega_1$. Three different 
2D shapes are considered for $\Omega_{1}$: circle, ellipse and bullet, all with the same area. This two-layer emulsion-based 
vesicle is surrounded by a surfactant layer (blue boundary) having a mass transfer coefficient $P$ (m/s).}  
\label{configurations}
\end{figure}

To fix ideas, without loss of generality, we consider $\Omega_0$ as a circle centered at the origin with radius $R_0=40\,\mu\text{m}$ (see Fig.~\ref{configurations}). On the other hand, the shape of $\Omega_1$ becomes a key factor of the release and we analyze the dependence of the drug delivery on the geometry of the shell. Three different shapes are considered for $\Omega_{1}$, each centered at the origin and of the same area (see Fig.~\ref{configurations}): 
\begin{itemize}
\item circle with radius $R_{1}$;
\item ellipse with horizontal semi-axis length $\gamma R_{1}$ and vertical semi-axis length $R_{1}/\gamma$ where $\gamma > 1$;
\item bullet-like shape with boundary described by the following quartic (so-called bean) curve: 
$$(\hat{R}_{1}-x)^{4} + (\hat{R}_{1}-x)^{2}y^{2} + y^{4} - 2\hat{R}_{1}(\hat{R}_{1}-x)[(\hat{R}_{1}-x)^{2}+y^{2}] = 0,$$
where choosing $\hat{R}_{1} = \sqrt{\frac{3\sqrt{3}}{7}} R_{1} \approx 0.8616 R_{1}$ ensures the area enclosed by the quartic curve is the same as the areas of the circle and ellipse above.
\end{itemize}
For all three configurations, the size of $\Omega_{1}$ is characterized by $R_{1}$ so we choose $\chi = R_{1}$ in the non-dimensionalization (\ref{eq:nondim_constants})--(\ref{gh6}). For the ellipse configuration, the eccentricity, $e$, is defined, which is related to $\gamma$ by the formula $e = \sqrt{1 - 1/\gamma^{4}}$. At initial time, we assume all the drug is in the internal core ($C_0=1$) while the external shell is assumed empty ($C_1=0$).

\begin{figure}[t]
\def\figureheight{0.2\textwidth}
\centering
\includegraphics[width=0.25\textwidth]{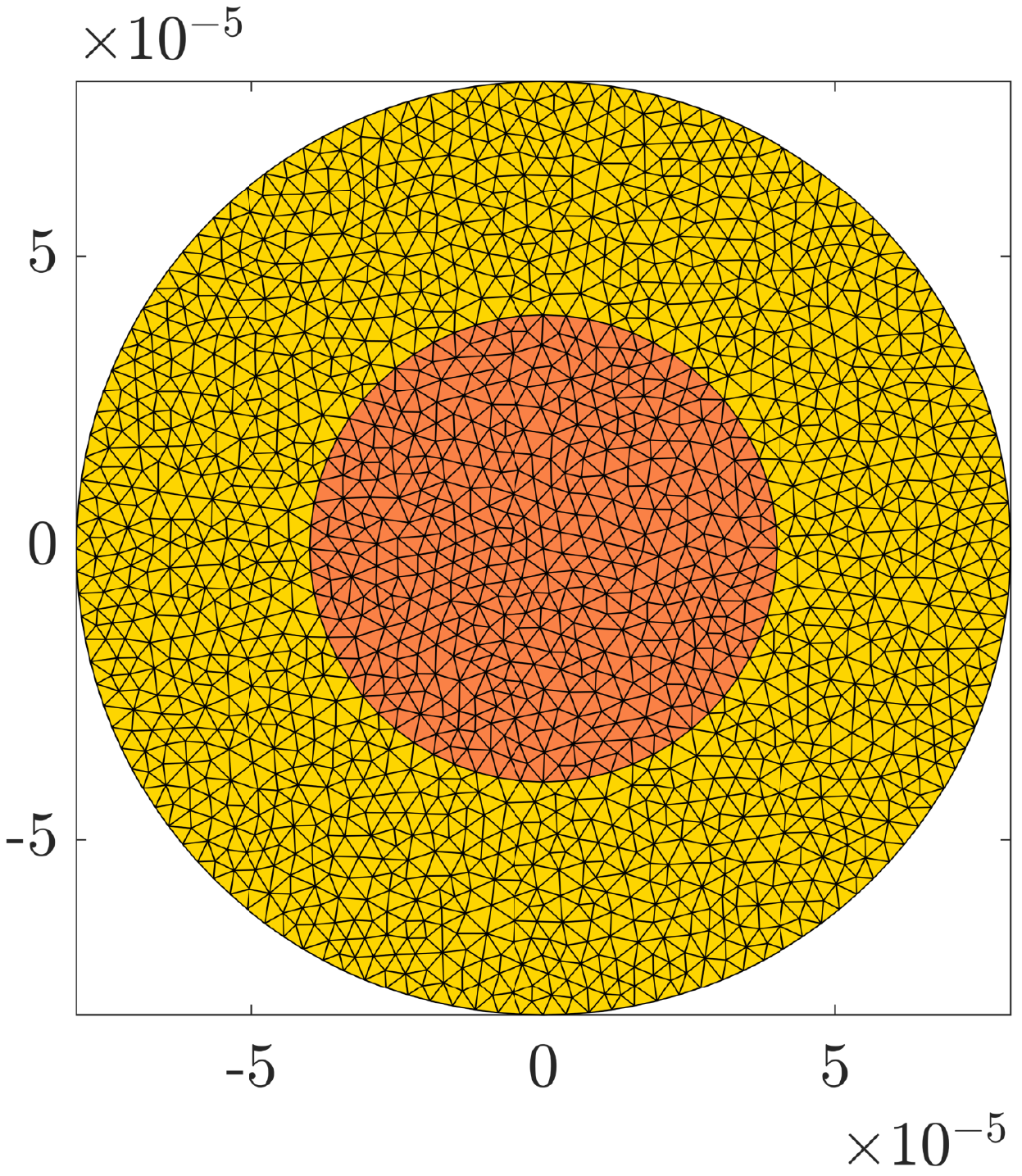}\hspace{0.03\textwidth}\includegraphics[width=0.45\textwidth]{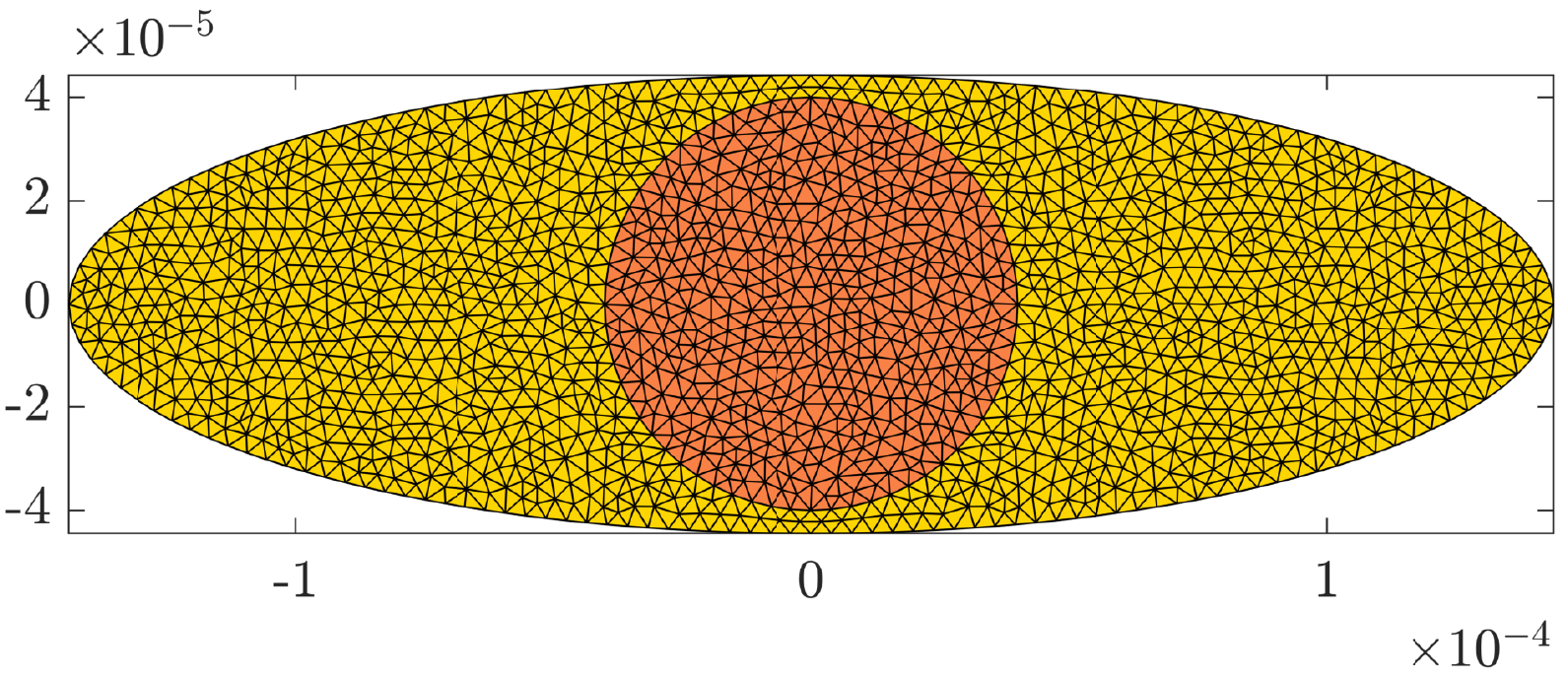}\hspace{0.03\textwidth}\includegraphics[width=0.2154\textwidth]{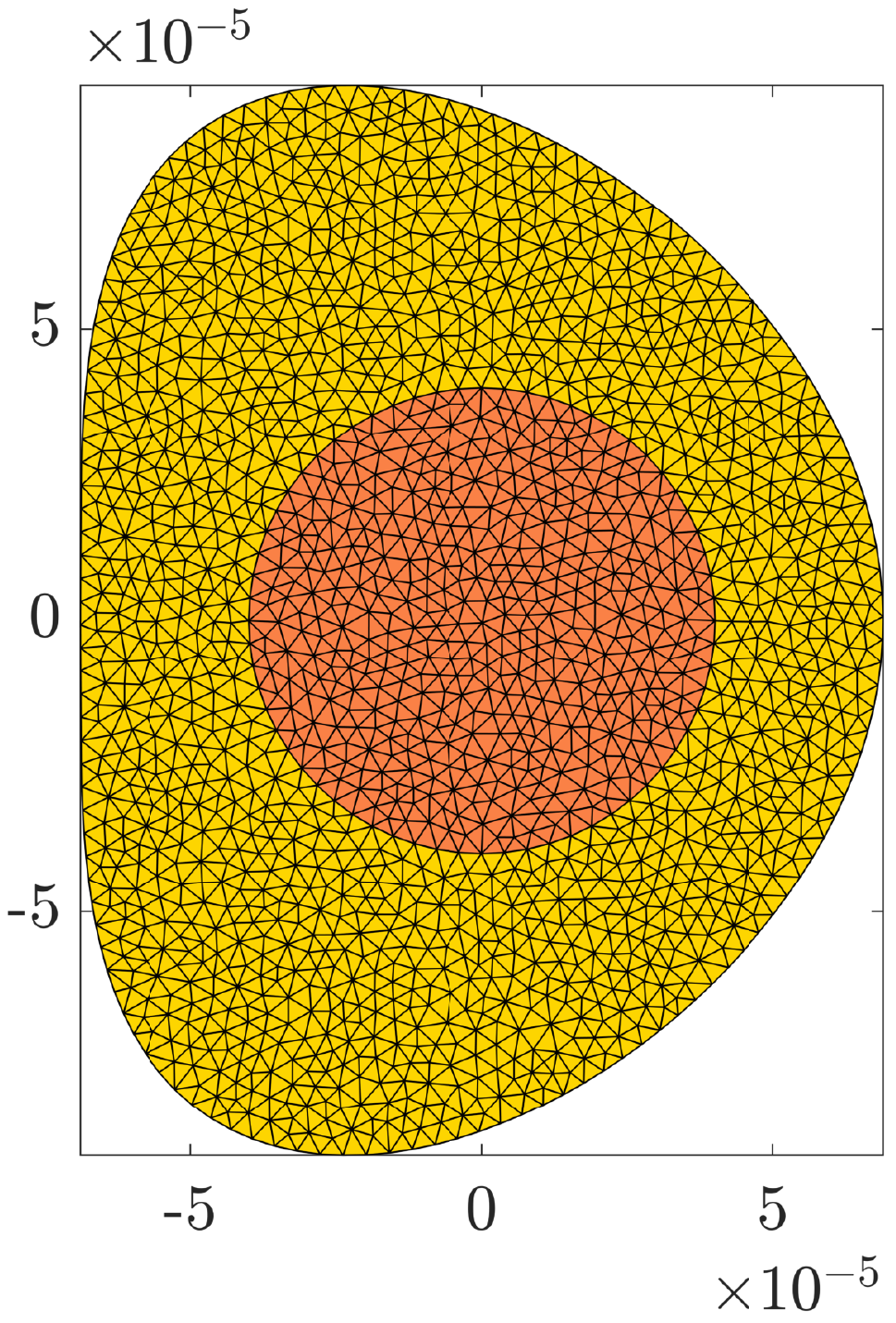}\hspace{0.2\textwidth}
\caption{Triangular meshes for the three droplet shapes of equal area: circle (left column) consisting of 1769 nodes and 3408 triangular elements, an ellipse with $e = 0.95$ (middle column) consisting of 1797 nodes and 3432 triangular elements and a bullet (right column) consisting of 1785 nodes and 3438 triangular elements.}
\label{fig:meshes}
\end{figure}

\begin{figure}[p]
\def\figureheight{0.2\textwidth}
\centering
\includegraphics[width=0.2375\textwidth]{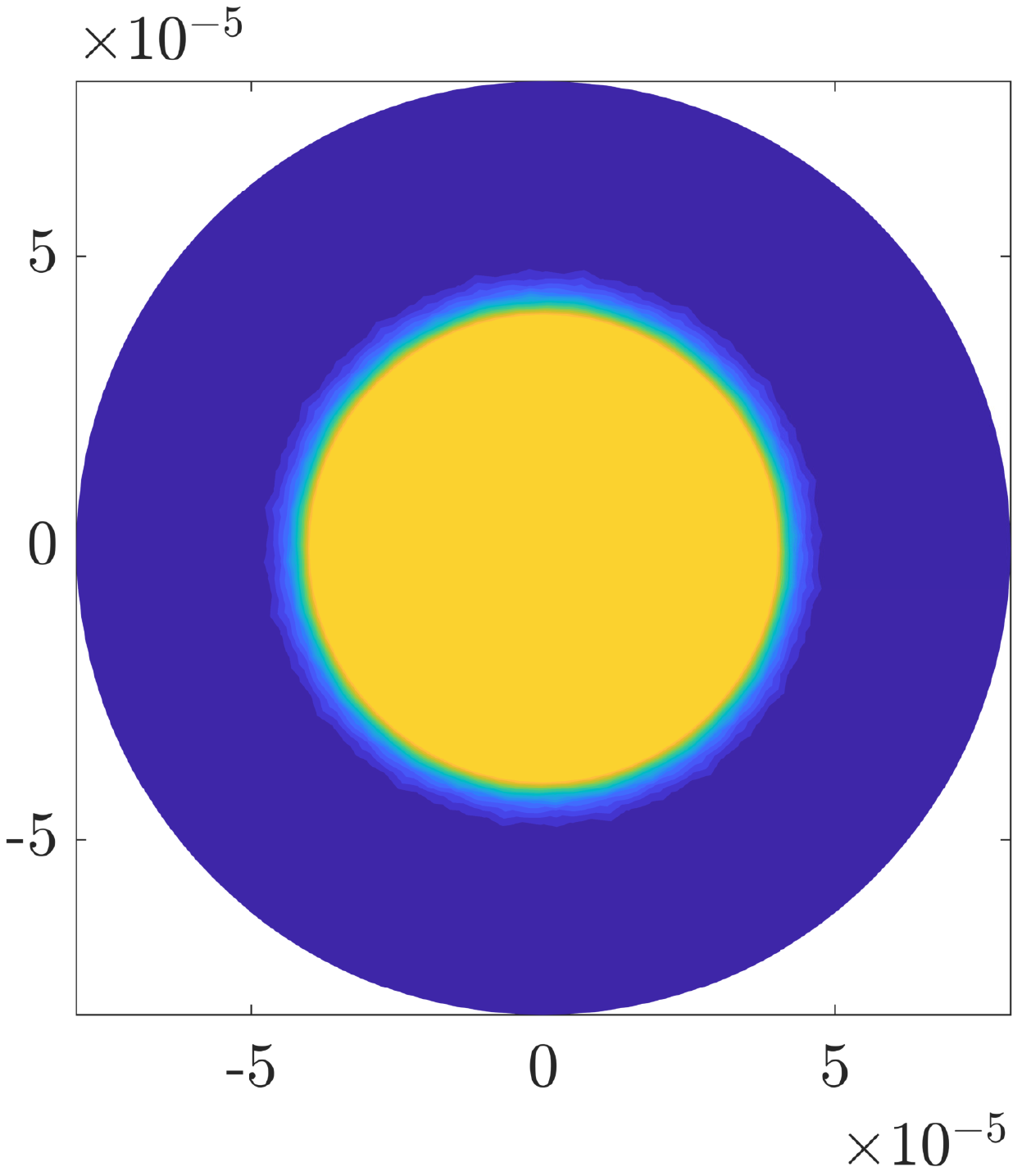}\hspace{0.03\textwidth}\includegraphics[width=0.4275\textwidth]{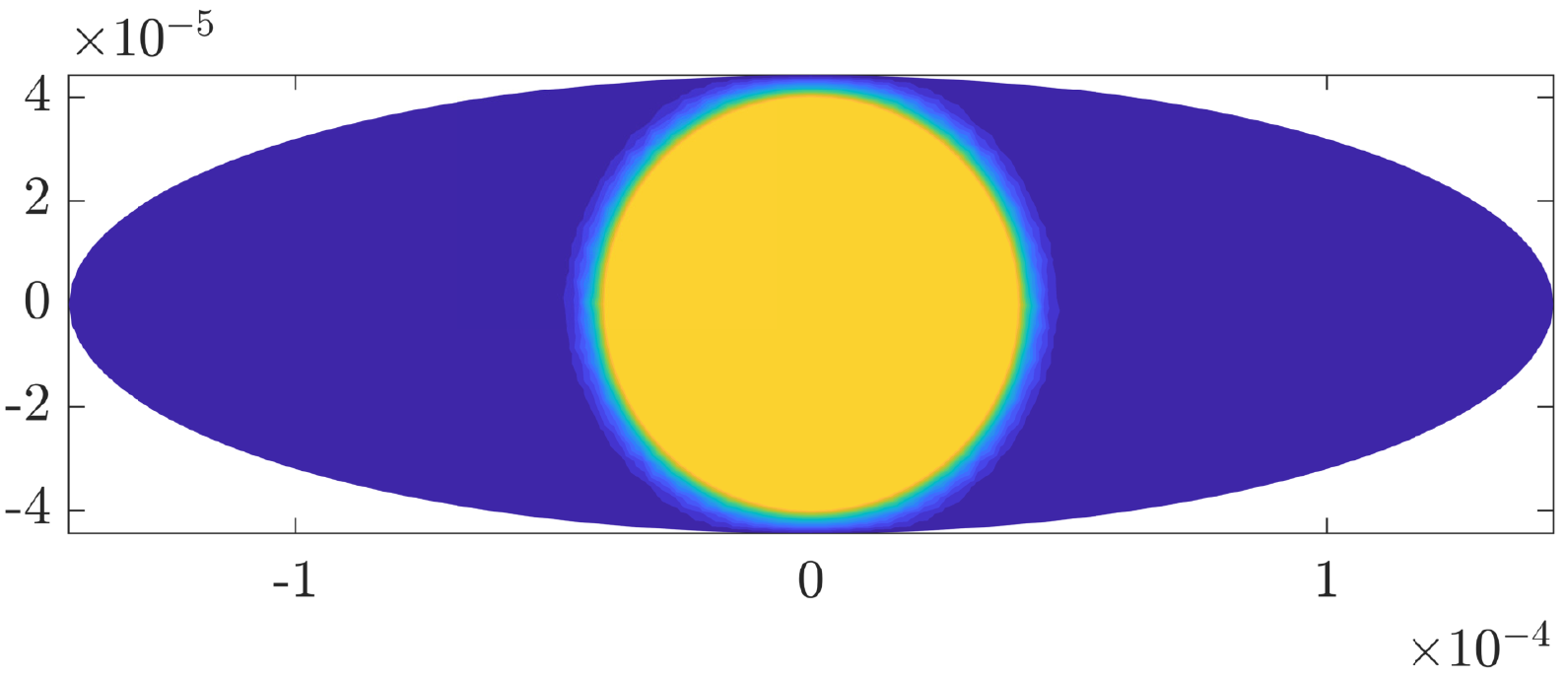}\hspace{0.03\textwidth}\includegraphics[width=0.2046\textwidth]{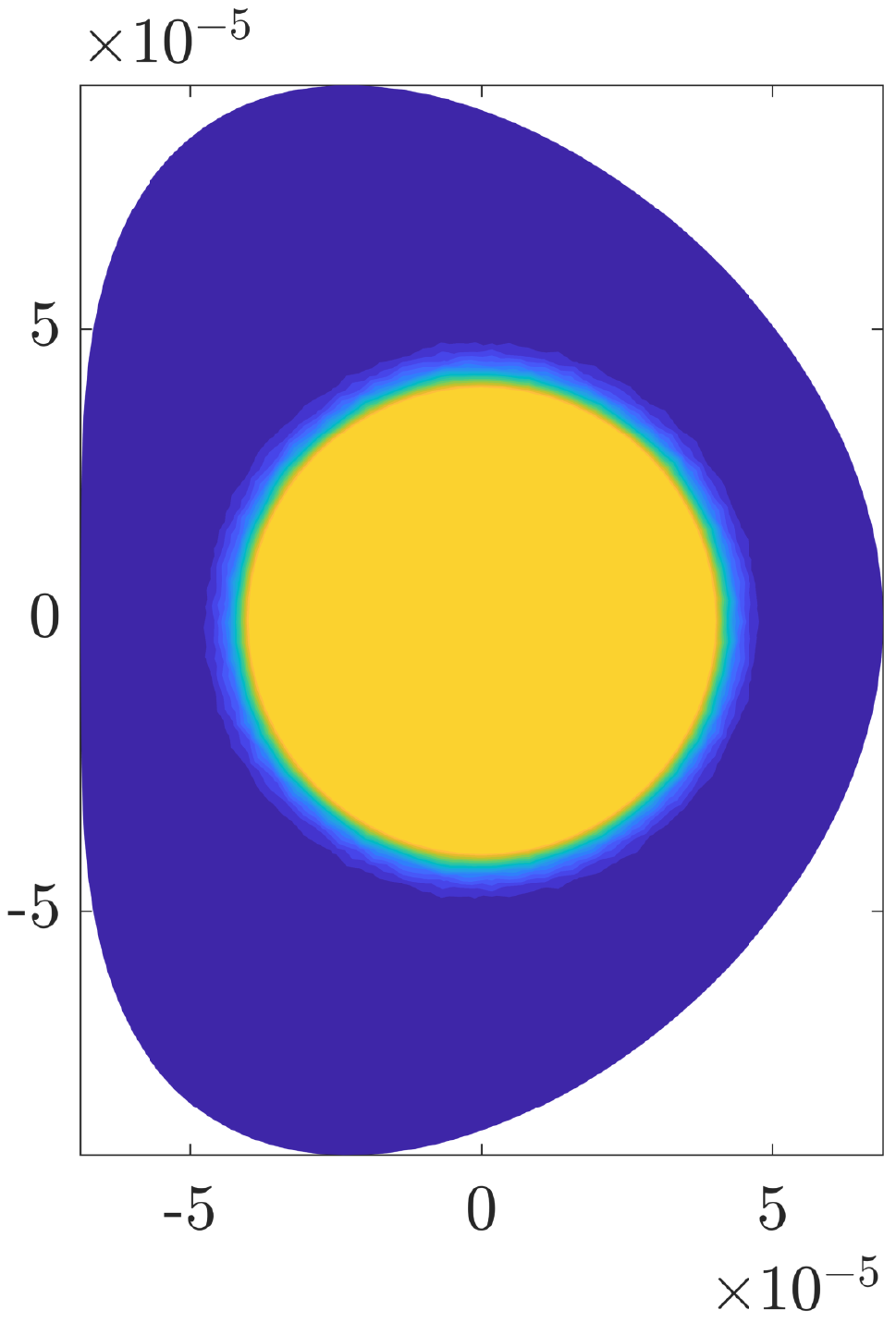}\hspace{0.03\textwidth}\includegraphics[height=0.2\textwidth]{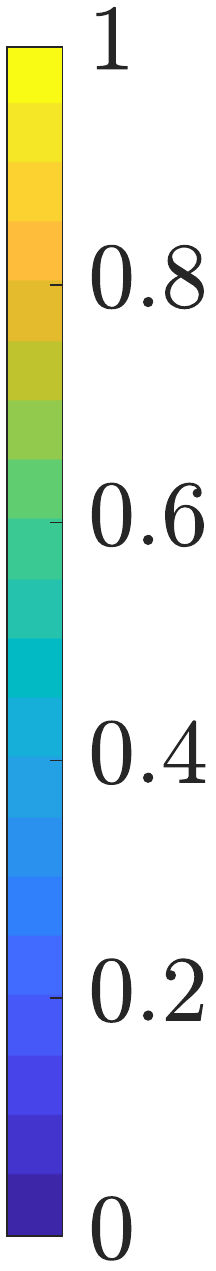}\\
\includegraphics[width=0.2375\textwidth]{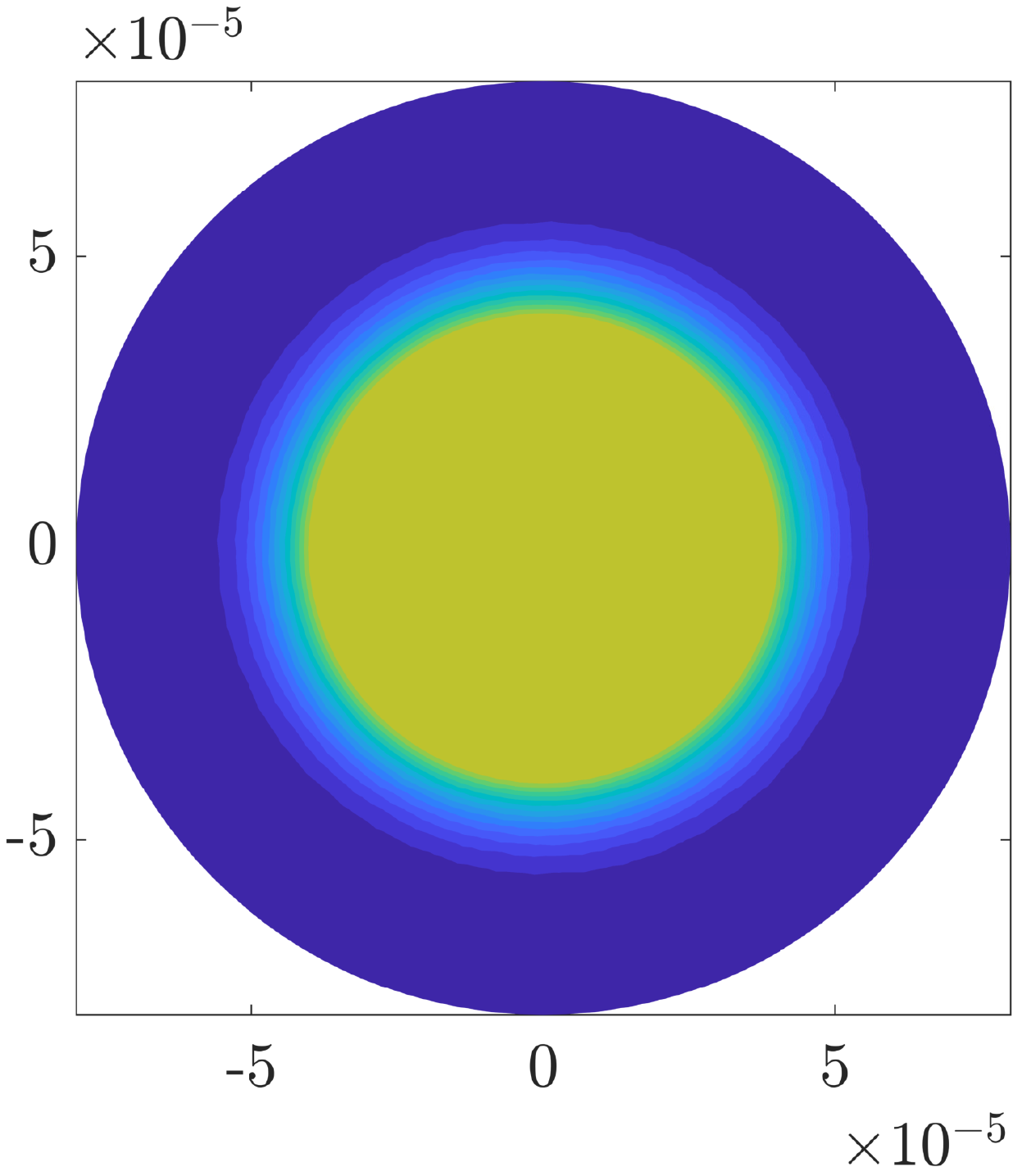}\hspace{0.03\textwidth}\includegraphics[width=0.4275\textwidth]{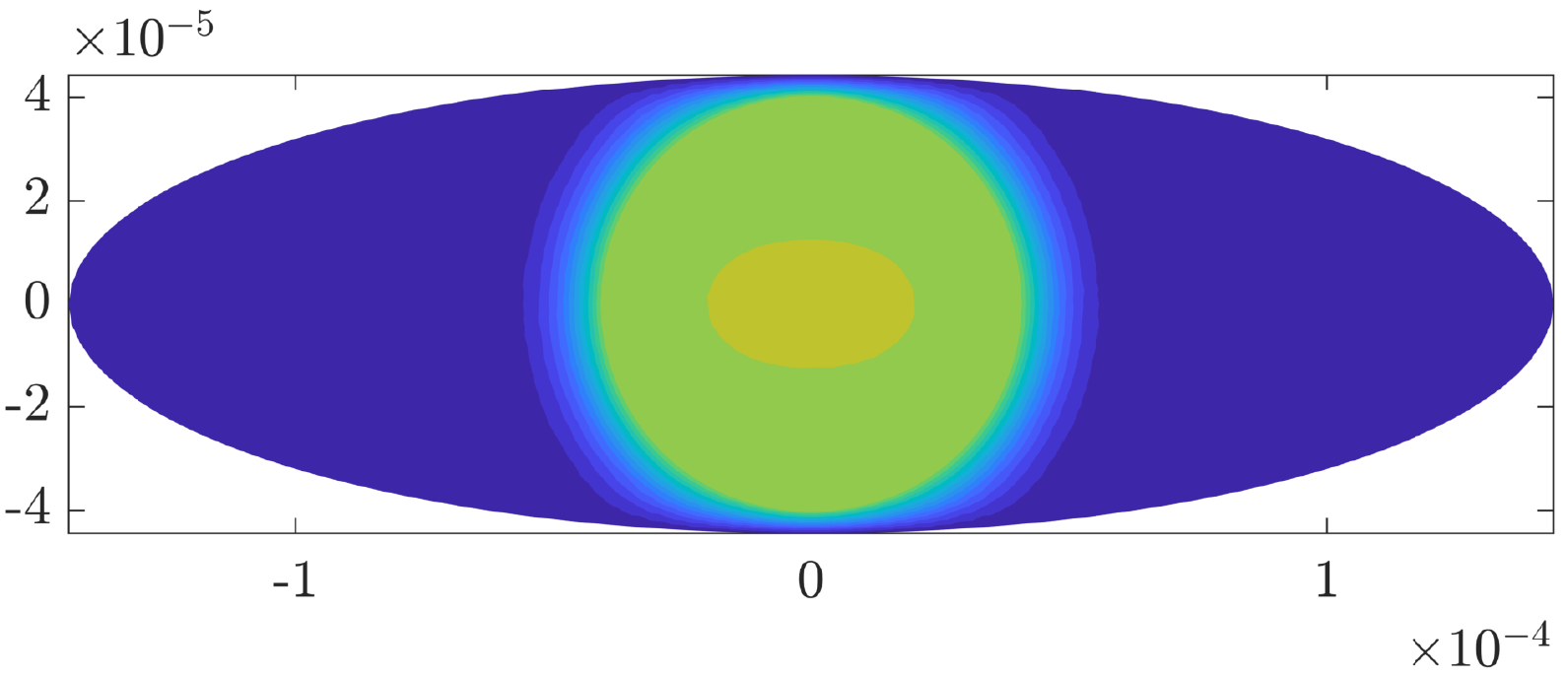}\hspace{0.03\textwidth}\includegraphics[width=0.2046\textwidth]{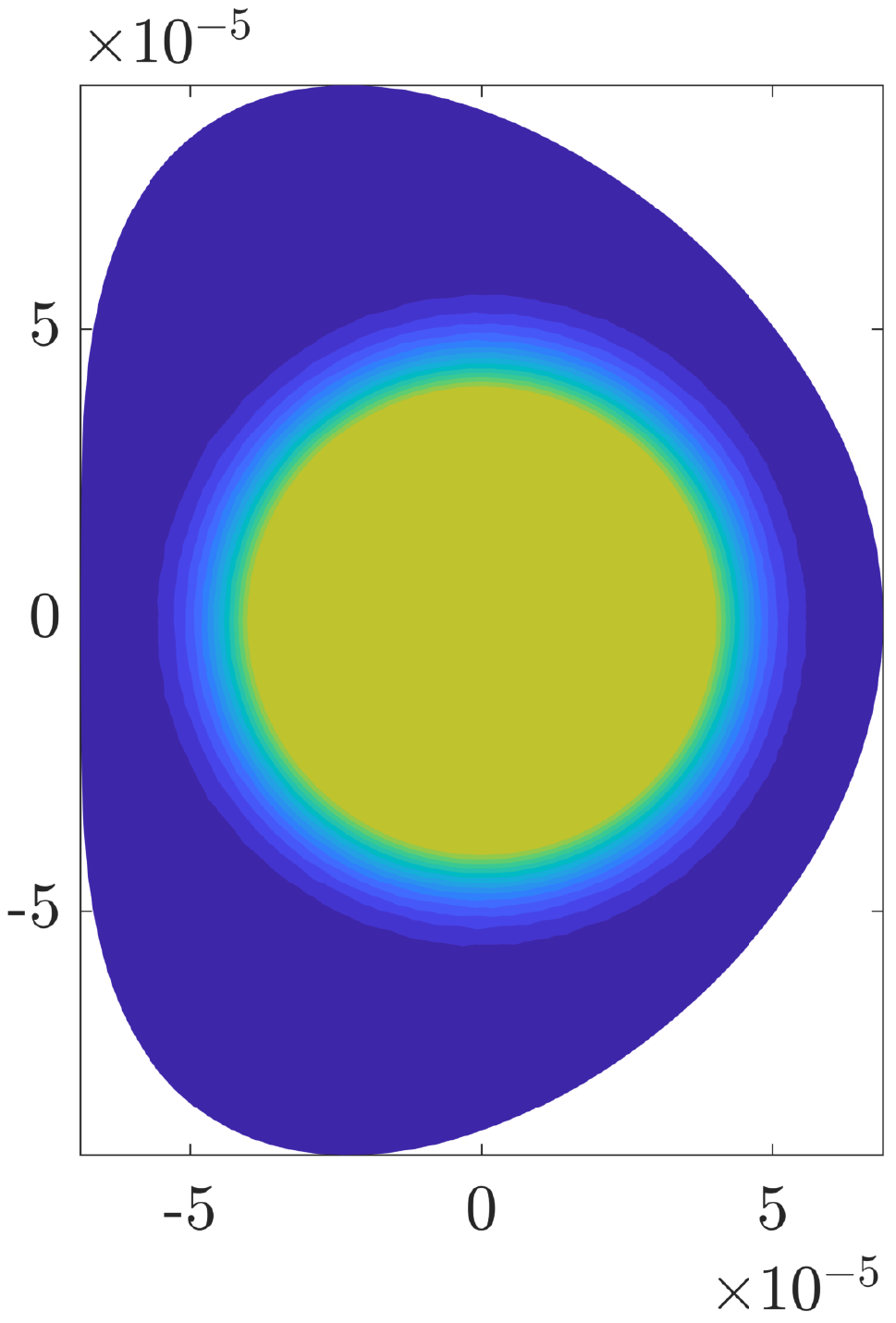}\hspace{0.03\textwidth}\includegraphics[height=0.2\textwidth]{colorbar.pdf}\\
\includegraphics[width=0.2375\textwidth]{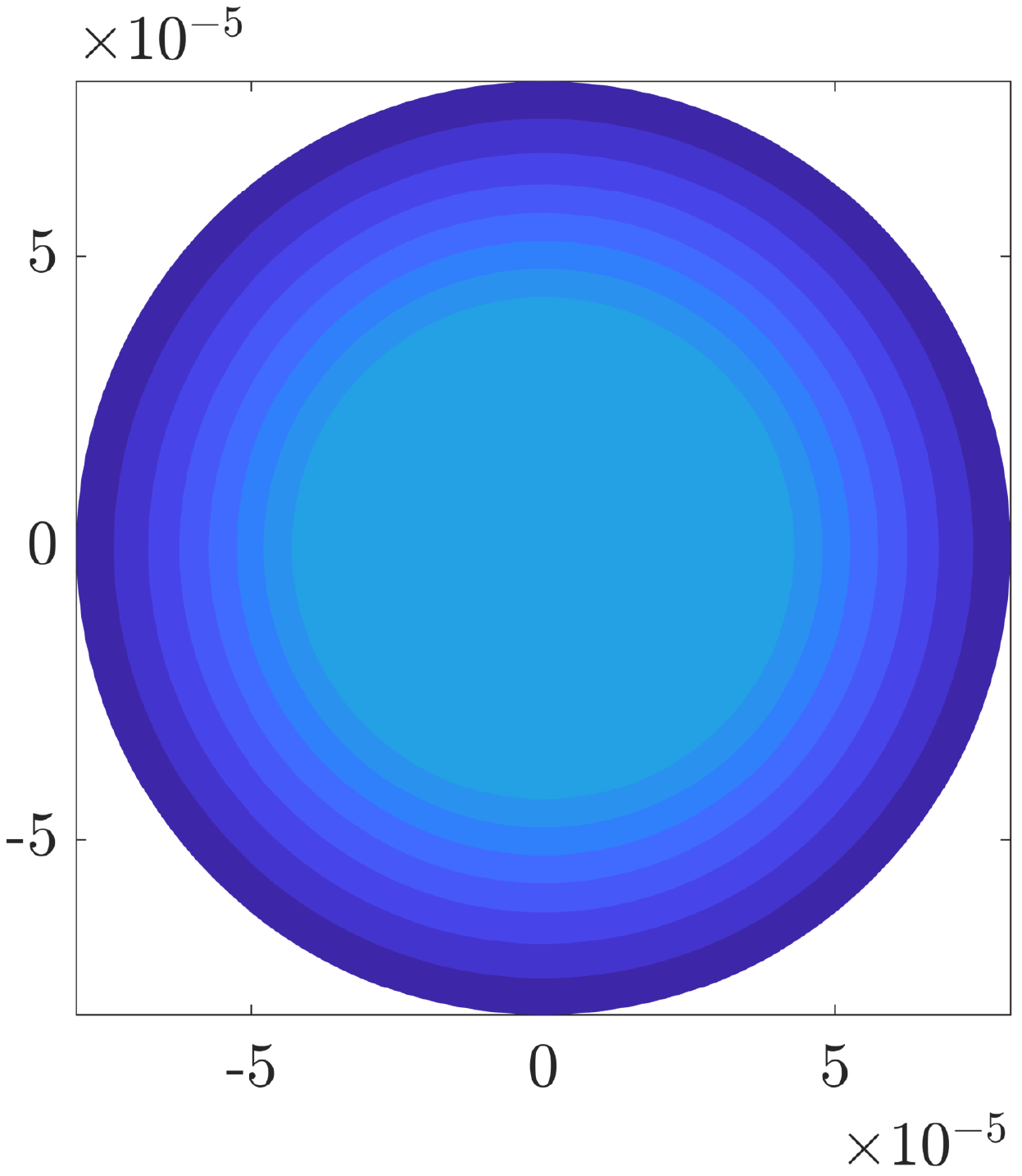}\hspace{0.03\textwidth}\includegraphics[width=0.4275\textwidth]{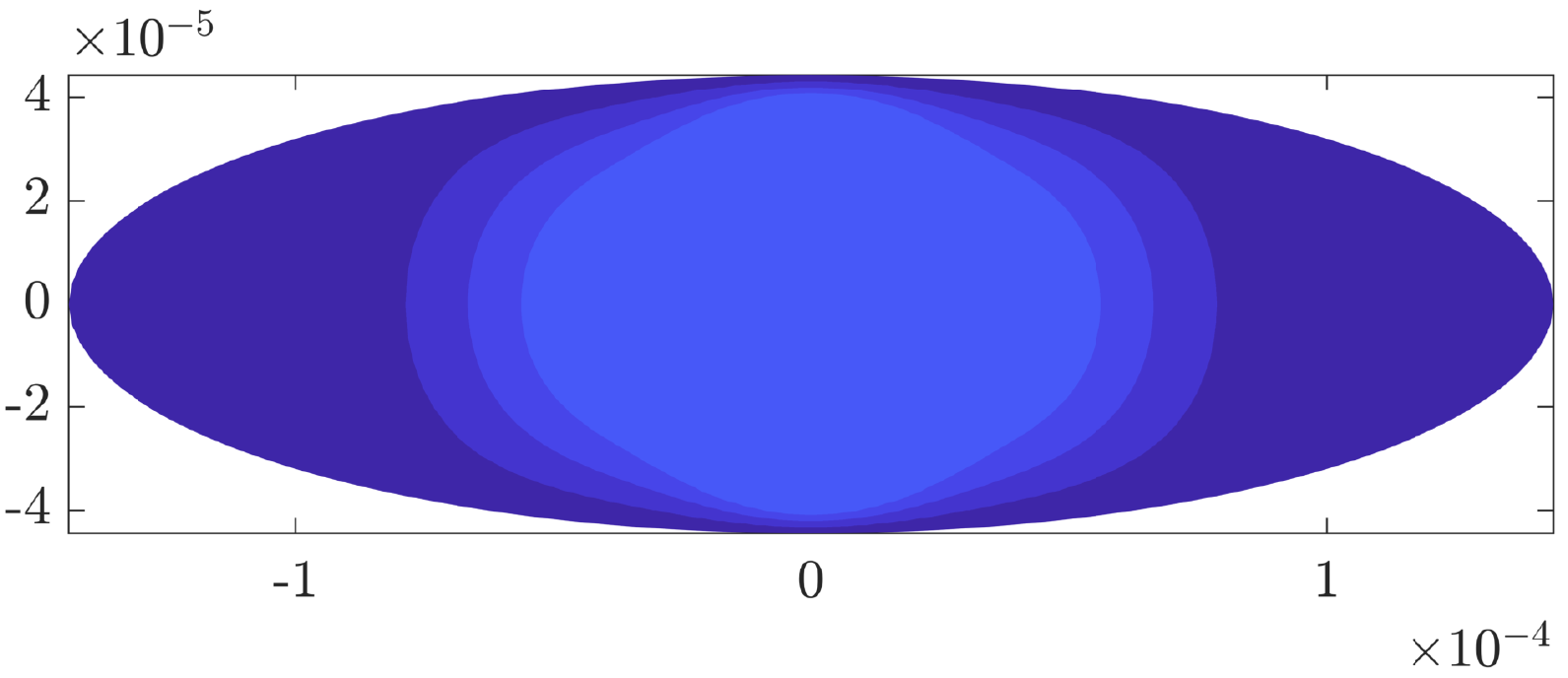}\hspace{0.03\textwidth}\includegraphics[width=0.2046\textwidth]{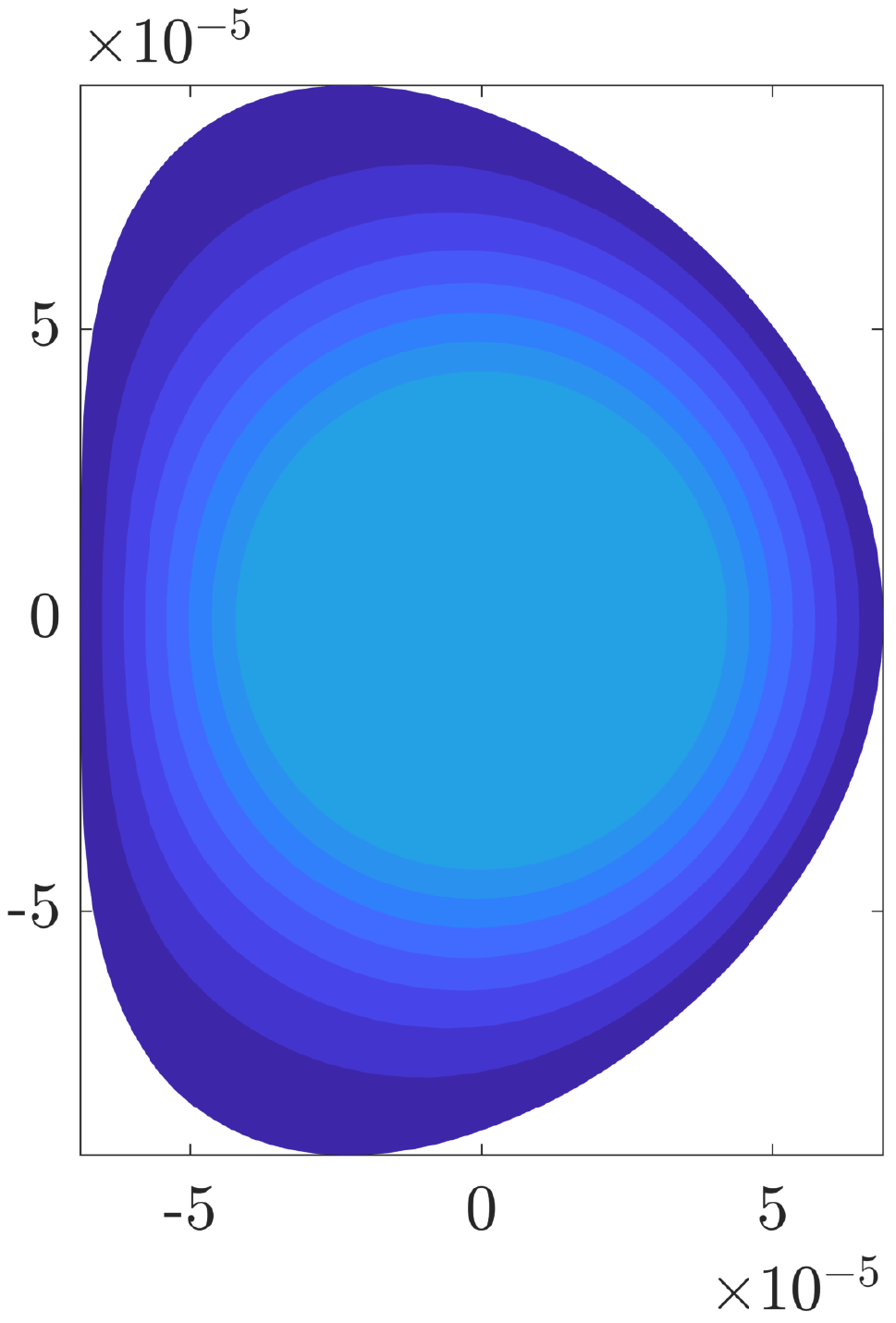}\hspace{0.03\textwidth}\includegraphics[height=0.2\textwidth]{colorbar.pdf}\\
\caption{Concentration field at $t = 1\,\text{min}$ (top row), $t = 6\,\text{mins}$ (middle row) and $t = 1\,\text{hr}$ (bottom row) for the three droplet shapes of equal area: circle (left column), an ellipse with $e = 0.95$ [$\gamma = 1.8$] (middle column) and a bullet (right column) (cfr. Fig.~\ref{fig:results_mass}). Results are produced using the meshes shown in Fig.~\ref{fig:meshes} and the parameter values: $R_0=40\,\mu\text{m}$, $R_1=80\,\mu\text{m}$, $D_1=10^{-13}\,\text{m}^2/\text{s}$, $P=2 \cdot 10^{-4}\,\text{m}/\text{s}$.}
\label{fig:results_concentration}
\end{figure}

First, we present results for the case of $R_{1} = 80\,\mu\text{m}$ and $\gamma = 1.8$. The finite volume discretisation outlined in Sect.~\ref{sec:numerical_concentration} is performed using the unstructured meshes shown in Fig.~\ref{fig:meshes}. Each of these meshes have an equivalent level of refinement with the prescribed mesh element size at all points used to describe the geometries in GMSH set to be equal\footnote{See GMSH documentation available at \href{http://gmsh.info/}{http://gmsh.info/} for more details.}. Further refining of the mesh did not visually alter the concentration fields (grid independence). Fig.~\ref{fig:results_concentration} shows the concentration field in the case of the different droplet configurations (left-right) at three times (top-down). It turns out that for the above parameters, the sensitive values are in the range: $10^{-8} \leq P \leq 10^{-3}$. In the limit, for $P<10^{-9},$ the surface acts as an impermeable barrier (as $P \rightarrow 0$) (release prevented), for $P>10^{-1},$  the droplet surface results in perfect contact with the surrounding external medium (as $P \rightarrow \infty$) (fastest release).

\begin{figure}[p]
\def\figureheight{0.24\textwidth}
\centering
\includegraphics[height=\figureheight]{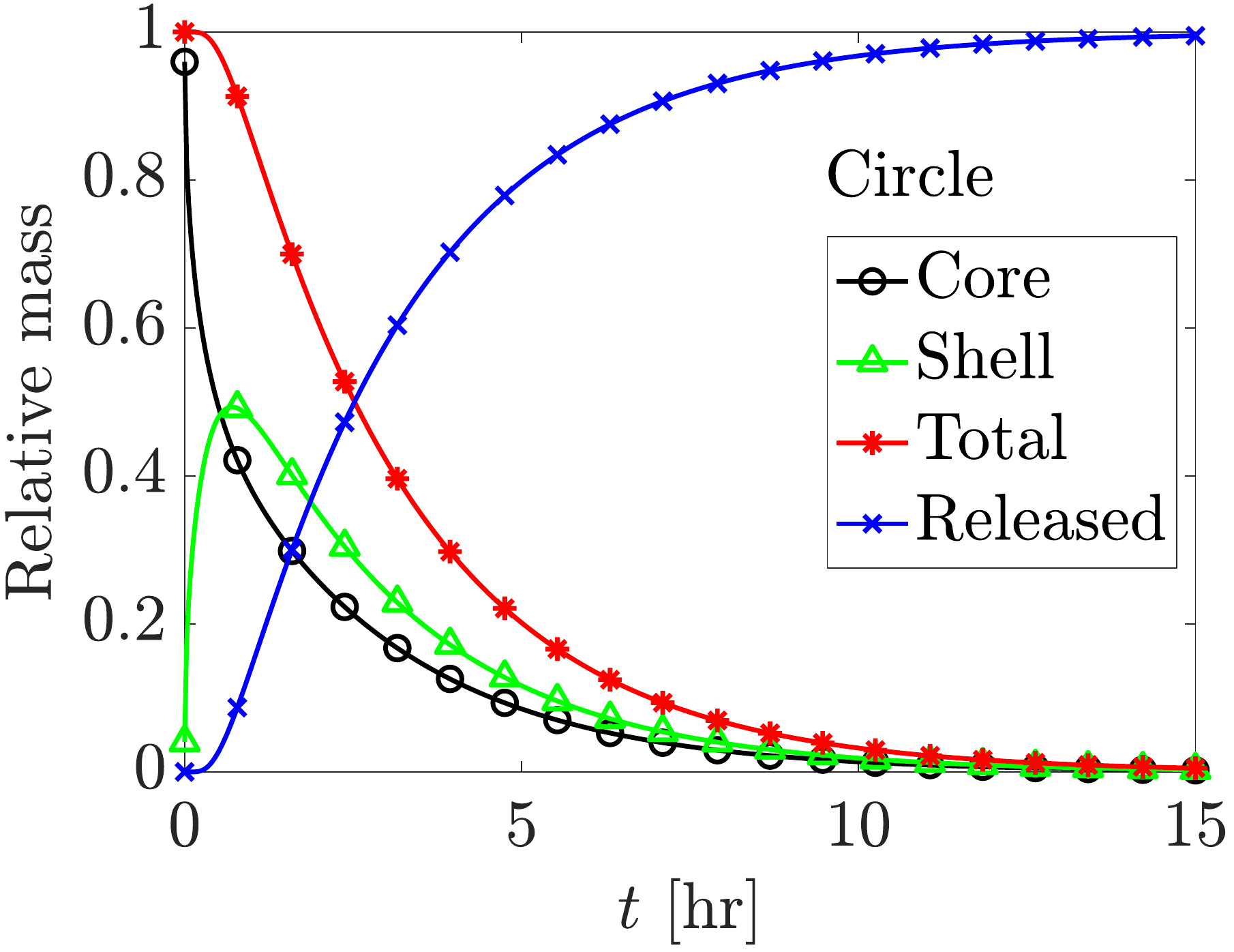} \hspace{0.01\textwidth}
\includegraphics[height=\figureheight]{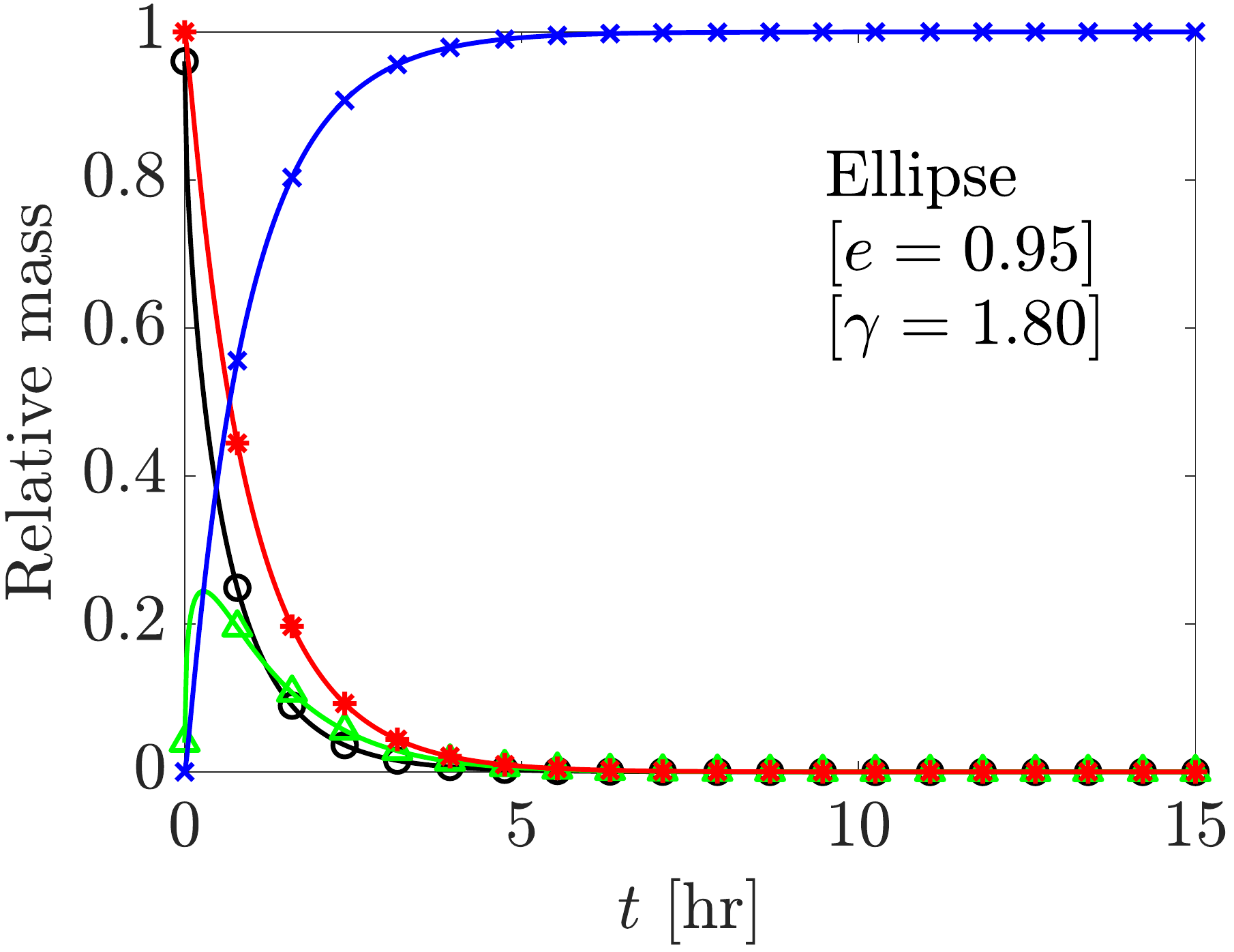} \hspace{0.01\textwidth}
\includegraphics[height=\figureheight]{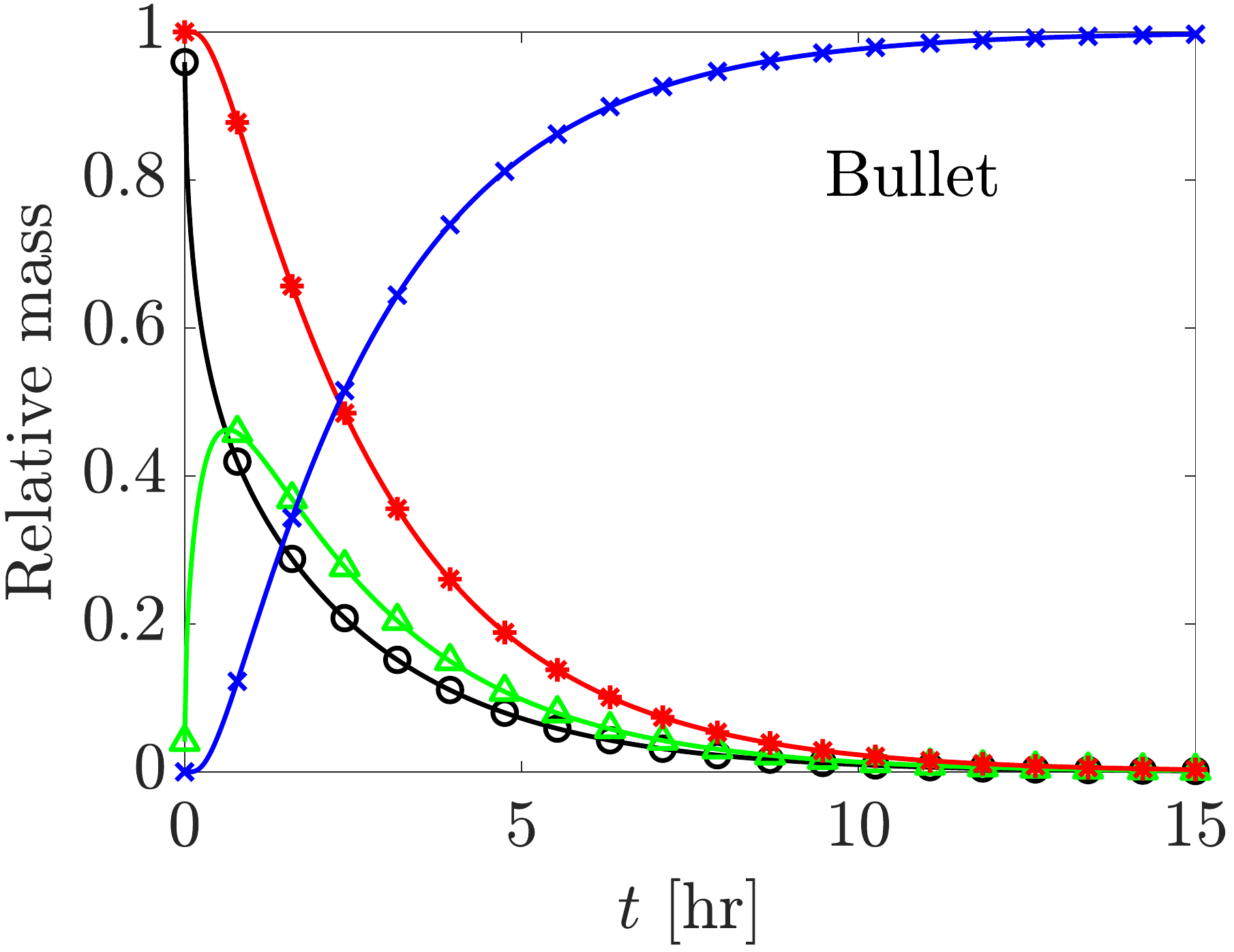}
\caption{Mass profiles for the three droplet shapes of equal area (cfr.~Fig.~\ref{fig:results_concentration}). $\text{Relative mass} = M_{0}(t)/M_{T}(0)$ (Core), $M_{1}(t)/M_{T}(0)$ (Shell), $M_{T}(t)/M_{T}(0)$ (Total) and $M_{r}(t)/M_{T}(0)$ (Released) [cf. Eqs (\ref{eq:M0_M1})--(\ref{eq:fractional_released_mass})]. Results are produced using the meshes shown in Fig.~\ref{fig:meshes} and the parameter values used in Fig.~\ref{fig:results_concentration}. Legend applies across all three figures.}
\label{fig:results_mass}
\end{figure} 

\begin{table}[p]
\renewcommand*{\arraystretch}{1.1}
  \begin{center}
\caption{Release time for a circular droplet for varying values of $R_1$ ($\mu$m), $D_1$ ($\text{m}^2/\text{s}$) and $P$ ($\text{m}/\text{s}$). All other parameters are held fixed: $R_0=40\,\mu\text{m}$ and $D_{0} = 10^{-10}\,\text{m}^2/\text{s}$.}
 \label{tab:release_time_circle}
\begin{tabular*}{0.6\textwidth}{@{\extracolsep{\fill}}lllrr}
\hline
$R_1$ &  $D_1$ & $P$  &  RT (hr) &  RT (HH:MM:SS) \\
\hline
$50 $& $ 10^{-12}$& $2 \cdot 10^{-4} $ & 0.41 & 00:24:38\\
$50 $& $ 10^{-12}$& $1 \cdot 10^{-7} $ & 0.87 & 00:51:57\\
$80 $& $ 10^{-12}$& $ 2 \cdot 10^{-4}  $ & 1.95 & 01:56:57\\
$ 80 $& $ 10^{-12}$& $ 1 \cdot 10^{-7}  $ & 2.59 & 02:35:36\\
$50$& $  10^{-13}$& $ 2 \cdot 10^{-4}$  & 4.08 & 04:04:34\\
$50$& $  10^{-13}$& $ 1 \cdot 10^{-7}$  & 4.51 & 04:30:42\\
$80$& $  10^{-13}$& $ 2 \cdot 10^{-4}$  & 19.47 & 19:28:20\\ 
$80$& $  10^{-13}$& $ 1 \cdot 10^{-7}$  & 20.08 & 20:04:42\\
\hline
\end{tabular*}
\end{center}
 \end{table}

\begin{table}[p]
\renewcommand*{\arraystretch}{1.0}
\begin{center}
\caption{Release time (\ref{eq:release_time}) from ellipsoidal droplets with the same area as a function of eccentricity. The values of eccentricity ($e$) are rounded to two decimal places and correspond to $\gamma = 1,1.2,\hdots,1.9$. We also compare with the circular shape ($e=0$) and the bullet shape. All other parameters are held fixed: $R_0=40\,\mu\text{m}$, $R_1=80\,\mu\text{m}$, $D_{0} = 10^{-10}\,\text{m}^2/\text{s}$, $D_1=10^{-13}\,\text{m}^2/\text{s}$ and $P=2 \cdot 10^{-4}\,\text{m}/\text{s}$.}
 \label{tab:release_time_ellipse}
\begin{tabular*}{0.6\textwidth}{@{\extracolsep{\fill}}llrr}
\hline
Shape & $e$ &  RT (hr) & RT (HH:MM:SS)\\   \hline\hline
Circle & 0 & 19.47 & 19:28:21\\\hline
Ellipse & 0.56 & 19.01 & 19:00:53\\
 & 0.72 & 17.79 & 17:47:14\\
 & 0.81 & 16.16 & 16:09:23\\
 & 0.86 & 14.33 & 14:19:46\\
 & 0.90 & 12.46 & 12:27:23\\		
 & 0.92 & 10.62 & 10:37:00\\
 & 0.94 & 8.86 & 08:51:18\\
 & 0.95 & 7.18 & 07:10:36\\
 & 0.96 & 5.55 & 05:32:52\\\hline
Bullet & N/A & 17.86 & 17:51:28\\
\hline				
\end{tabular*}
\end{center}
\end{table}

\begin{figure}[p]
\def\figureheight{0.36\textwidth}
\centering
\includegraphics[height=\figureheight]{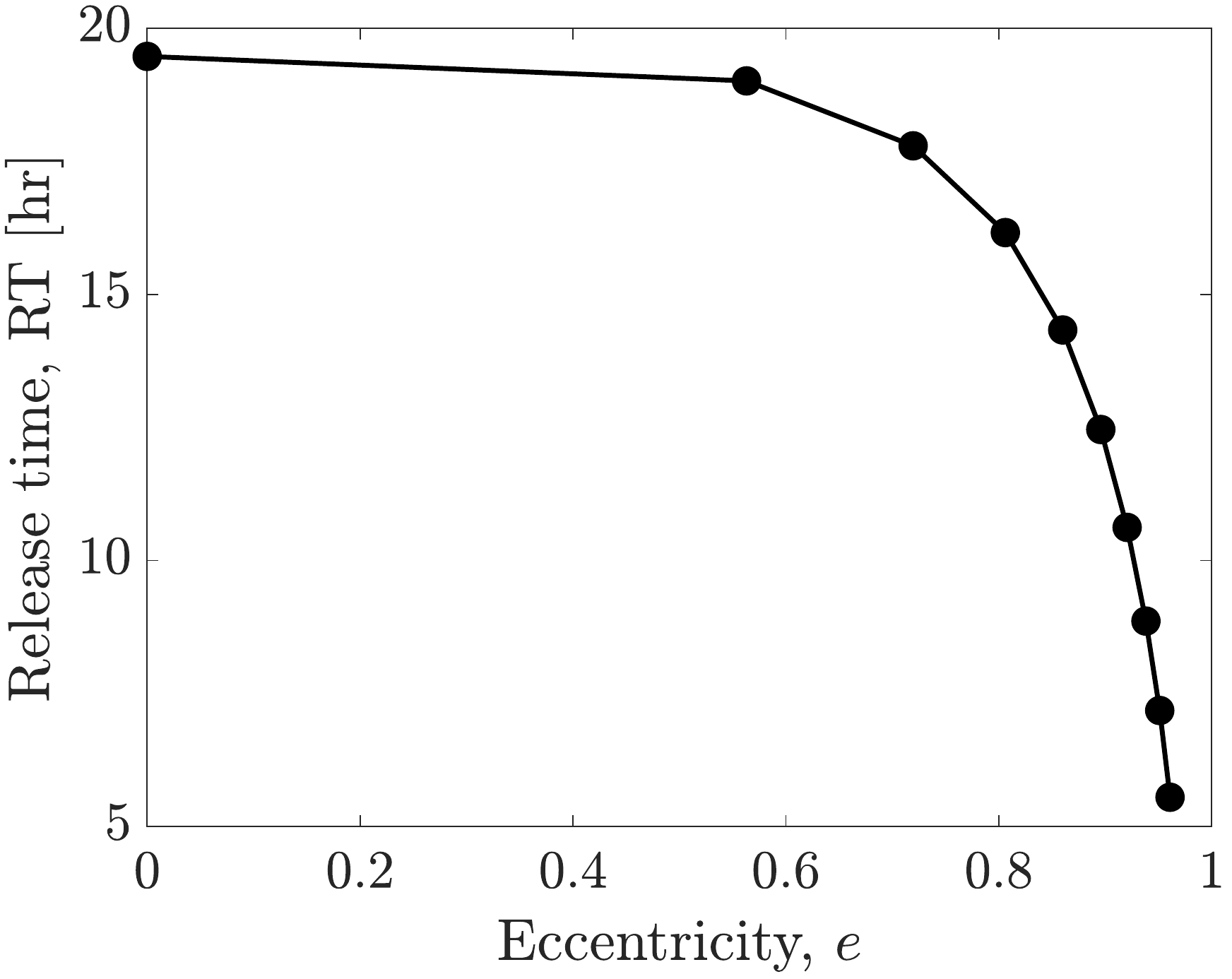} 
\caption{Reduction of the release time (RT, Eq. (\ref{eq:release_time})) for the ellipsoidal shaped droplet with increasing eccentricity ($e$) (cfr. Table \ref{tab:release_time_ellipse}). All other parameters are held fixed: $R_0=40\,\mu\text{m}$, $R_1=80\,\mu\text{m}$, $D_{0} = 10^{-10}\,\text{m}^2/\text{s}$, $D_1=10^{-13}\,\text{m}^2/\text{s}$ and $P=2 \cdot 10^{-4}\,\text{m}/\text{s}$.}
\label{fig:results_eccentricity}
\end{figure}  

\begin{figure}
\def\figureheight{0.2\textwidth}
\centering
\includegraphics[width=0.2375\textwidth]{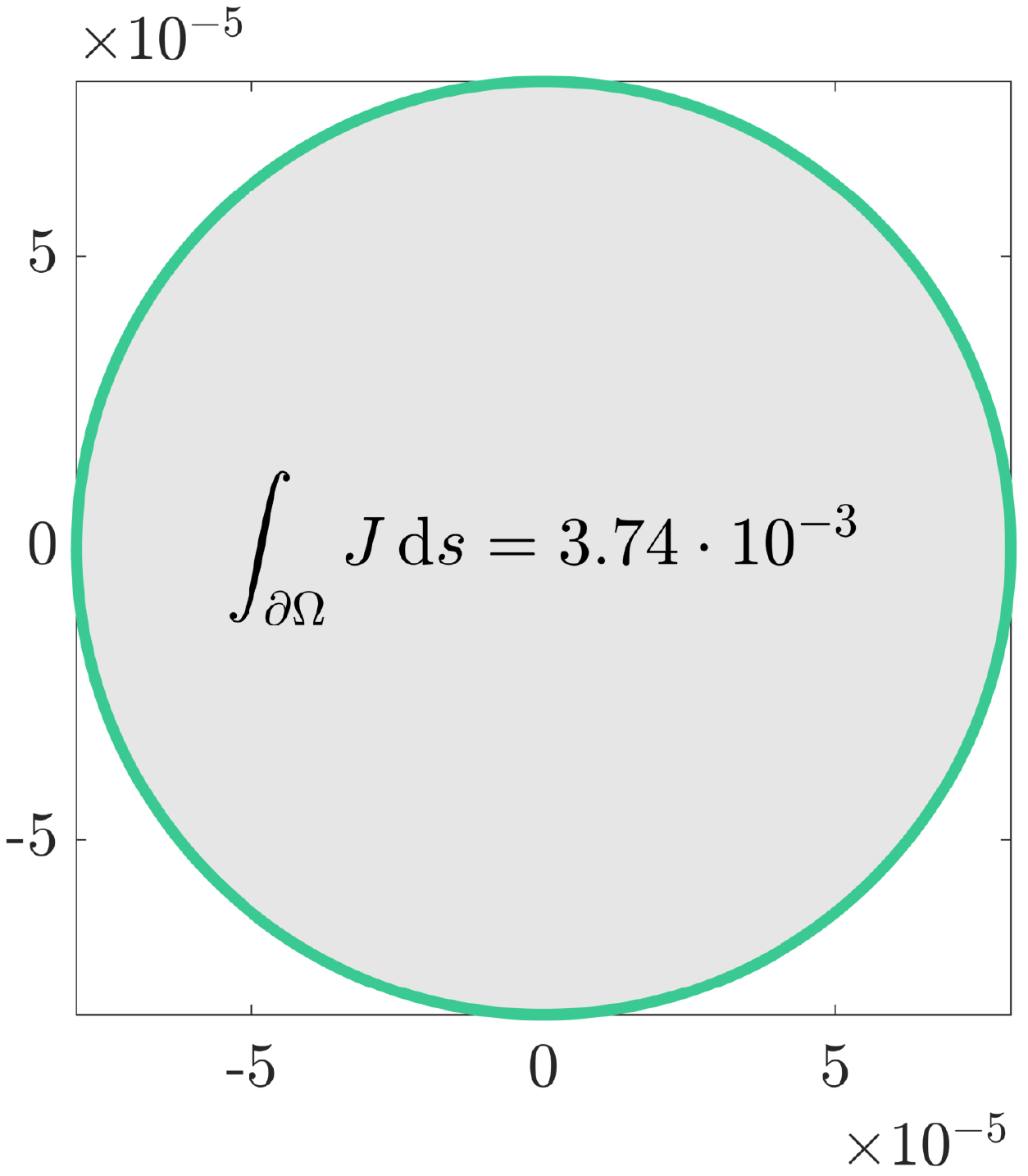}\hspace{0.03\textwidth}\includegraphics[width=0.4275\textwidth]{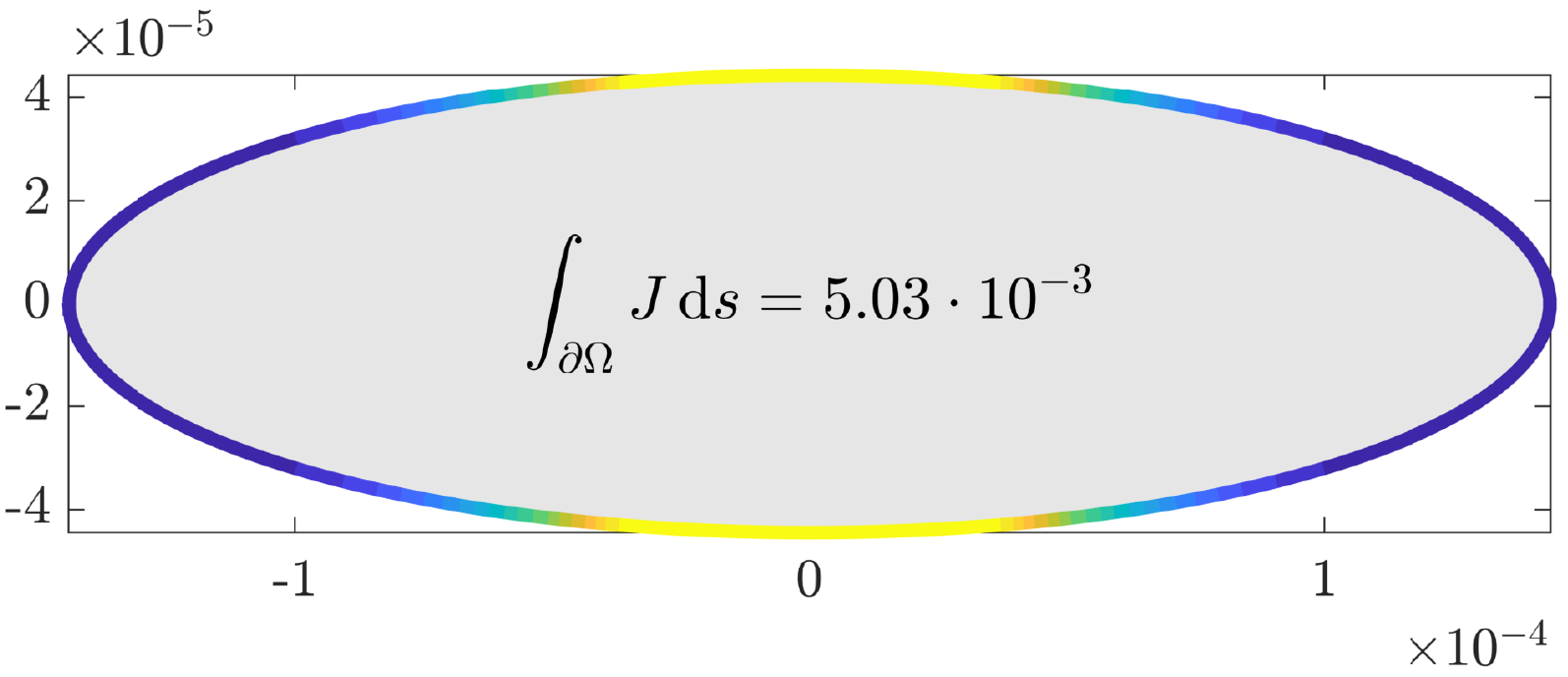}\hspace{0.03\textwidth}\includegraphics[width=0.2046\textwidth]{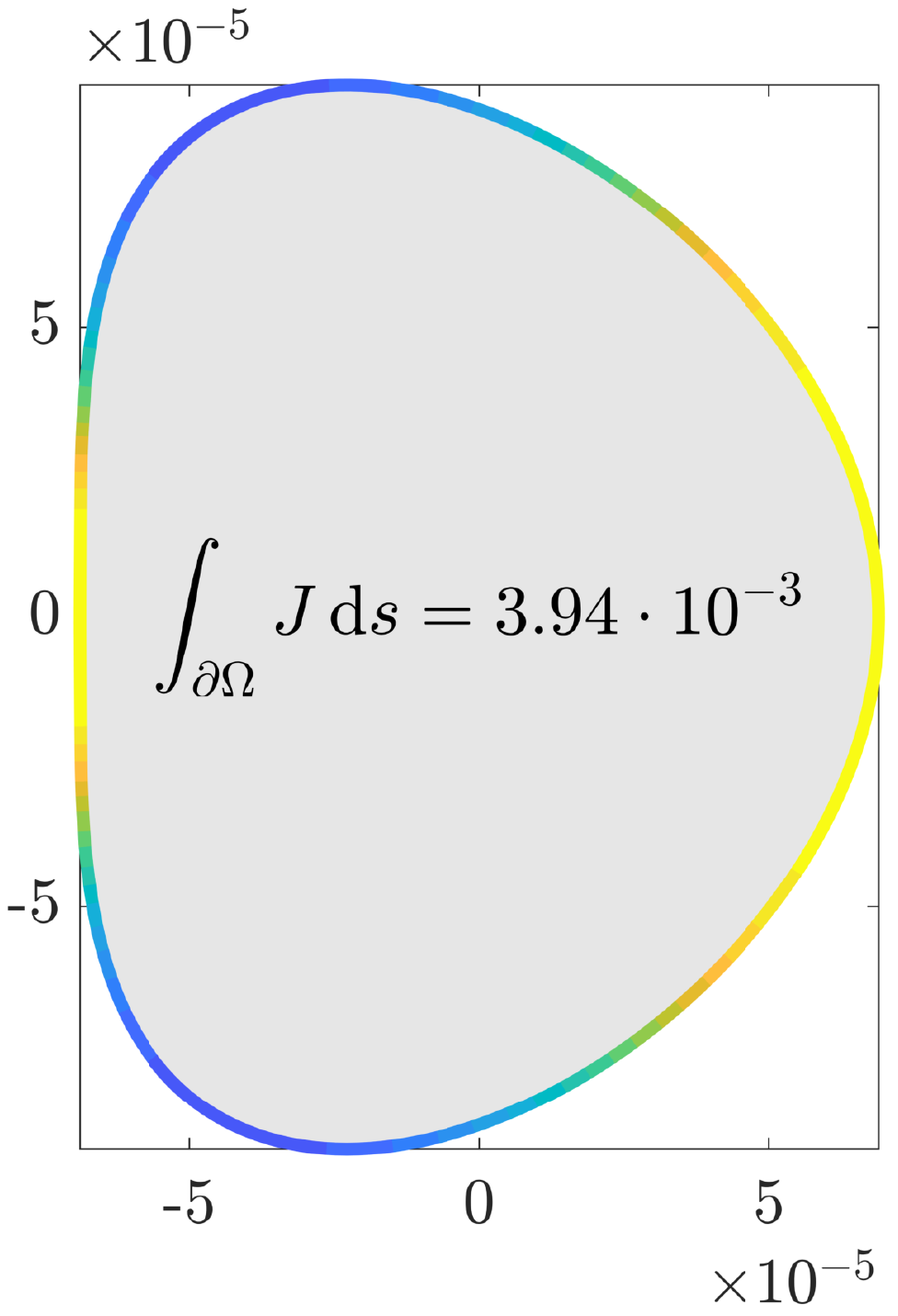}\hspace{0.03\textwidth}\includegraphics[height=0.18\textwidth]{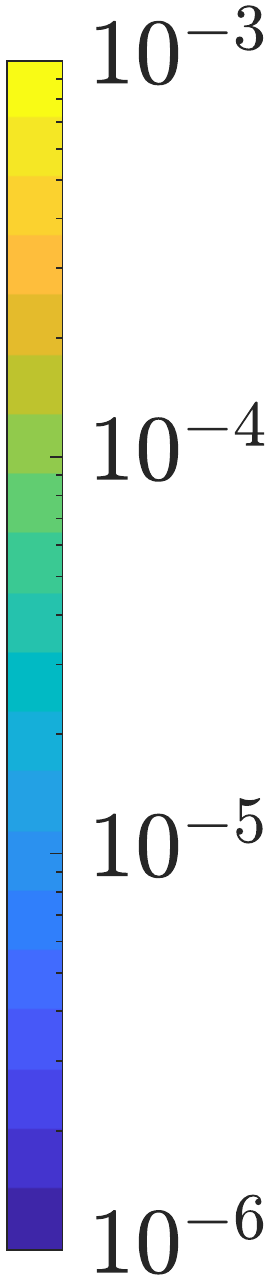}\\
\caption{Boundary flux $J = -D_{1}\nabla c_{1}\cdot \boldsymbol{n}_{\Omega}$ at $\boldsymbol{x}\in\partial\Omega$ and total boundary flux $\int_{\partial\Omega} J\,\text{d}s$ at $t = 1\,\text{hr}$ for the three droplet shapes of equal area: circle (left column), an ellipse with $e = 0.95$ [$\gamma = 1.8$] (middle column) and a bullet (right column) (cfr. Fig.~\ref{fig:results_concentration}, bottom row). Results are produced using the meshes shown in Fig.~\ref{fig:meshes} and the parameter values used in Fig.~\ref{fig:results_concentration}.}
\label{fig:results_bflux}
\end{figure}

The effect of the combined multi-layer diffusivity is similar to that of other releasing  systems \cite{dem2010}. Drug mass is transported from the core to the surrounding shell, and thereafter released to the external medium. Mass is monotonically decreasing in the core, but is first increasing up to some upper bound and then decaying asymptotically in the shell layer (Fig.~\ref{fig:results_mass}).  In the external release medium the mass progressively accumulates at a time depending on the diffusive properties of the two-layer droplet and the resistance $P$. In other words, due to the absorbing condition (\ref{eq:model_int2}), all drug mass is transferred to the surrounding environment at a sufficiently long time and the total mass is preserved and equals the initial value. 

A crucial indicator is the release time (RT), measured here as:
\be
\label{eq:release_time}
RT=\min \{ t  \,| \, M_r(t) \geq 0.999\}.
\ee

Additional simulations demonstrate that the time and the size of the mass peak in layer 1 is much more correlated with the diffusivity and the size of the shell, and much less controlled by the mass resistance of the surfactant $P$ (see Table \ref{tab:release_time_circle}). A more sustained release occurs in the case of a surfactant having a smaller mass transfer coefficient ($P = 10^{-7}$). In Table \ref{tab:release_time_ellipse} and Fig.~\ref{fig:results_eccentricity}, we analyze the dependence of the release time from the geometry (ellipses with different eccentricity and bullet-like shape) of the droplet, when the area of the vehicles remains the same.  It turns out that the bullet shape droplet has a RT comparable with that of the ellipse with $e=0.72$, and shorter of that of the circle.
Fig. \ref{fig:results_bflux} shows the
increased flux $J=-D_1 \nabla c_1 \cdot \boldsymbol{n}_{\Omega}$ (coloured contour) at the surface due to
the higher gradient of concentration,  in correspondence to points of lower curvature.
Moreover, the global flux $ \int_{\p \Omega}  J ds  $ exhibits a faster release rate for the ellipsoidal droplets.
In summary, all these results demonstrate that an oblate shape promotes a faster drug delivery, while a round geometry guarantees a more sustained release.

\section{Conclusions} 

Multiple emulsions are highly structured fluids consisting of drops that encapsulate smaller droplets inside. The availability of such multi-compartment vesicles with controlled size and structure have attracted much attention as robust and versatile drug delivery systems, in equilibrium with the external flow. In this work we analyze the structure of a double emulsion in which each drop contains a single internal droplet, thus developing  a core-shell structure whose core diameter and shell thickness and shape can be controlled. A two-layer diffusion model for the drug release is developed and solved numerically. Results show the importance of the parameters on the drug kinetics, demonstrating how the oblate shape exhibits a faster drug delivery, while a round geometry promotes a more sustained release. 
Additional efforts are needed to improve microfluidic platforms to generate and analyze fluid droplets with higher stability and biocompatibility and to achieve the successful translation of emulsion-based drug delivery systems into clinical applications. Therefore, in a future work, we plan to couple the present model with the microfluidics allowing the investigation of the effects of interaction of underlying flow and drug release under conditions mimicking the in-vitro and in-vivo systems. 
The predictive capability of the model will provide important guidance in fabricating double emulsions that can guarantee a controlled drug  delivery to the target sites at desired rates and time.

\section*{Acknowledgments}
GP, AT and SS acknowledge funding from the European Research Council under the European Unions Horizon 2020 Framework Programme (No.~FP/2014-2020)/ERC Grant Agreement No.~739964 (COPMAT).
\newpage
\bibliographystyle{plainnat}

\end{document}